\documentclass[11pt]{article}

\usepackage[dvipsnames]{xcolor}
\usepackage{hyperref}

\usepackage[T1]{fontenc}

\usepackage[defblank]{paralist}

\usepackage{latexsym}
\usepackage{stmaryrd}
\usepackage{amsmath,amsfonts}

\usepackage{mathtools}

\usepackage{xspace}
\usepackage{url}
\usepackage{graphicx}
\usepackage{complexity}
\usepackage{cite}
\usepackage{amssymb}

 \input{diagxy}
 \xyoption{curve}

\newtheorem{theorem}{Theorem}
\newtheorem{corollary}{Corollary}
\newtheorem{lemma}{Lemma}
\newtheorem{definition}{Definition}
\newtheorem{example}{Example}
\newtheorem{proposition}{Proposition}
\newtheorem{remark}{Remark}
\newtheorem{proof}{Proof}
\newtheorem{appthm}{A:}

\renewcommand{\vec}{\bar}

\newcommand{\set}[1]{\{ #1 \} \xspace}

\newcommand{\Hom}{\mathsf{Hom}}

\newcommand{\rcncovers}{\mathrel{\overset{\makebox[0pt]{\mbox{\normalfont\tiny RCN}}}{\rightrightarrows}}}
\newcommand{\Core}{\mathsf{Core}}

\newcommand{\homset}[2]{\ensuremath{\Hom(#1,#2)}}

\newcommand{\tupleletter}[1]{\ensuremath{\mathbf{#1}}}

\renewcommand{\a}{\tupleletter{a}}

\renewcommand{\cc}{\tupleletter{c}}

\newcommand{\x}{\tupleletter{x}}

\newcommand{\X}{\tupleletter{X}}
\newcommand{\Y}{\tupleletter{Y}}
\newcommand{\Z}{\tupleletter{Z}}

\newcommand{\BF}[1]{\ensuremath{\mathbf{#1}}}

\newcommand{\lscott}{[\![}
\newcommand{\rscott}{]\!]}
\newcommand{\sem}[1]{\ensuremath{\lscott  {#1} \rscott}}

\newcommand{\Rep}[2][\mathsf{OCWA}^*]{\ensuremath{\sem{#2}_{\mathsf{#1}}}}

\newcommand{\canrep}[1]{\ensuremath{\chi_p{(#1)}}}

\newcommand{\multicore}[1]{\ensuremath{\varpi (#1)}}

\newcommand{\cterm}[2]{\ensuremath{\{ {#1} \mathbin{|} {#2} \}}}

\newcommand{\Max}[2]{\ensuremath{\operatorname{Max}_{#1}(#2)}}

\newcommand{\homo}[3]{\ensuremath{{#1} :  #2 \rightarrow #3}}

\newcommand{\comp}{\circ}

\usepackage{fullpage}
\usepackage{authblk}

\title{On Equivalence and Cores for\\
 Incomplete Databases in Open and Closed Worlds\\
(Technical Report)} 

\author[1,2]{Henrik Forssell}
\author[3,1]{Evgeny Kharlamov}
\author[1]{Evgenij Thorstensen}

\affil[1]{University of Oslo, Norway} 
\affil[2]{University of South East Norway}
\affil[3]{Bosch Center for Artificial Intelligence, Germany}
\affil[ ]{\textit {\{jonf,evgeny.kharlamov,evgenit\}@ifi.uio.no}}
\affil[ ]{\textit {evgeny.kharlamov@de.bosch.com}}

\date{}

\begin{document}

\maketitle

\begin{abstract}
Data exchange heavily relies on the notion of incomplete database instances. Several semantics for such instances have been proposed and include open (OWA), closed (CWA), and open-closed (OCWA) world. For all these semantics important questions are: whether one incomplete instance semantically implies another; when two are semantically equivalent; and whether a smaller or smallest  semantically equivalent instance exists. For OWA and CWA these questions are fully answered. For several variants of OCWA, however, they remain open. In this work we adress these questions for Closed Powerset semantics and the OCWA semantics of \cite{DBLP:journals/jcss/LibkinS11}. We define a new OCWA semantics, called OCWA*, in terms of homomorphic covers that subsumes both semantics, and characterize semantic implication and equivalence in terms of such covers. This characterization yields a guess-and-check algorithm to decide equivalence, and shows that the problem is NP-complete. For the minimization problem we show that for several common notions of minimality there is in general no unique minimal equivalent instance for Closed Powerset semantics, and consequently not for the more expressive OCWA* either. However, for Closed Powerset semantics we show that one can find, for any incomplete database, a unique finite set of its subinstances which are subinstances (up to renaming of nulls) of all instances semantically equivalent to the original incomplete one. We study properties of this set, and extend the analysis to OCWA*.
\end{abstract}

\section{Introduction}
\label{sec:motivation}

\paragraph*{Data Exchange}
Data exchange is the problem 
	of translating information
	structured under a \emph{source schema} 
	into a \emph{target schema}, 
	given a source data set and a set of declarative \emph{schema mappings} 
	between the source and target schemata.
This problem has originally been studied for traditional relational databases 
where a decade of intensive research brought up a number of foundational and system oriented work~\cite{DBLP:books/cu/ArenasBLM2014,DBLP:conf/sigmod/BernsteinM07,DBLP:conf/pods/Kolaitis05,DBLP:conf/birthday/FaginHHMPV09,DBLP:conf/vldb/PopaVMHF02}.
More recently research in data exchange changed its focus in various directions that include non-relational~\cite{DBLP:journals/jacm/ArenasL08} and temporal data~\cite{DBLP:journals/is/GolshanaraC20}, knowledge bases~\cite{DBLP:journals/ai/ArenasBCR16}, mapping discovery~\cite{DBLP:conf/pods/CateK0T18,DBLP:journals/tods/CateKQT17},
and probabilistic settings~\cite{DBLP:journals/tkde/KimmigMMG19,DBLP:conf/aaai/LukasiewiczMPS16}.

In relational data exchange, 
	a set of schema mappings $M$ 
	is defined as a set of source-to-target tuple 
	generating dependences~\cite{DBLP:books/aw/AbiteboulHV95} of the form
		$\phi(\vec x, \vec y) \rightarrow \psi(\vec x, \vec z)$,
where $\phi(\vec x, \vec y)$ (resp.~$\psi(\vec x, \vec z)$) is a query over the source (resp.~target) schema with its variables.
In general such mappings only partially specify how to populate attributes of the target schema 
	with data from a given \emph{source  instance}~\cite{DBLP:journals/tcs/FaginKMP05},
	i.e., due to existential variables $\vec z$ in $\psi(\vec x, \vec z)$.
Therefore, 
	data exchange can result in possibly multiple incomplete 
	\emph{target  instances} $A$.
Each such $A$ represents a set of possible complete target instances
and there are several options on how such correspondence, or \emph{semantics of incomplete instances}, can be defined, including 
 Open World (OWA)~\cite{DBLP:journals/tcs/FaginKMP05,DBLP:journals/tods/FaginKP05}, Closed World (CWA)~\cite{DBLP:journals/tods/HernichLS11}, Open and Closed World (OCWA, with annotated instances)~\cite{DBLP:journals/jcss/LibkinS11}, and Powerset Closed World (PCWA)~\cite{Gheerbrant:2014:NEQ:2691190.2691194}, which we discuss in detail in  Section~\ref{sec:prelims}.

\paragraph*{Problems for Data Exchange}
In the context of data exchange the following questions have attracted considerable attention: 
given a semantics for incomplete database instances,  decide:
\begin{itemize}
	\item\textbf{[Semantic Implication:]} whether one incomplete instance semantically implies another; 
	\item \textbf{[Equivalence:]} whether two incomplete instances are semantically equivalent; and 
	\item \textbf{[Minimality, Core:]} whether a smaller or smallest (core) semantically equivalent incomplete instance exists. 
\end{itemize}
These questions form a natural progression, in that a characterization of semantic implication leads to one for equivalence, which in turn allows the study of minimal equivalent instances. 
The latter is important since, e.g., in some cases one can use the smallest minimal instance for computing certain answers by naively' evaluating queries directly on this instance. 

\paragraph*{How These Problems Have Been Addressed So Far}
These three questions are the focus of this paper since they have only partially been answered.
Indeed, for OWA and CWA, these questions have been fully answered. For OWA, semantic implication corresponds to the existence of a database homomorphism from one instance into another, and a unique smallest equivalent instance (the core~\cite{DBLP:journals/tods/FaginKP05}) always exists, and is minimal for several natural notions of minimality. Likewise, for CWA semantic implication corresponds to the existence of a strongly surjective homomorphism from one instance to another \cite{DBLP:journals/tods/HernichLS11}. This implies that equivalence corresponds to isomorphism, rendering the question of smallest equivalent instance moot.
For PCWA, semantic implication corresponds to the existence of a \emph{homomorphic cover} from one instance to another \cite{Gheerbrant:2014:NEQ:2691190.2691194}, while the question of smallest equivalent instance remains open. 
For OCWA with annotated instances, both questions are open, although preliminary results were previously presented by the authors~\cite{DBLP:conf/amw/ForssellKT17}.
Finally, we are not aware that the question of semantic implication between PCWA and OCWA with annotated instances has previously been considered. 

\paragraph*{Our Approach to Implication and Equivalence}
Therefore, 
in this paper we address the questions of Semantic Implication, Equivalence, and Minimality
for PCWA and OCWA semantics. 
To this end we introduce a novel open-and-closed-world semantics, OCWA*, based purely on the notion of homomorphic cover. 
We show how both PCWA and OCWA semantics with annotated instances can be defined as special cases of OCWA*.
This subsumption property allows us to characterize semantic implication and equivalence for all three semantics using homomorphic covers, and thus also semantic implication and equivalence between PCWA and OCWA with annotated instances. 

\paragraph*{Our Approach to Minimality and Cores}
We study several  natural notions of minimality, and show for all of them that there is in general no unique minimal equivalent instance for PCWA
nor, consequently, for the more expressive  OCWA*.
%
This raises the question: 
\emph{How can one find a smaller or `better'  equivalent instance?} 
Indeed,
even if one can find all equivalent subinstances of a given incomplete instance $A$
and compare them using the characterization of equivalence,  one still does not know whether a there exists a smaller equivalent instance that is not a subinstance of $A$.

We address this challenge as follows.
Focusing first on 
 PCWA,  we show  that for all instances $A$ there exists a  finite set \multicore{A} of `PCWA-cores'  which serves to determine all minimal instances that are equivalent to $A$. 
 More precisely, this set has the following properties: 
\begin{enumerate}
	\item  each member of \multicore{A}  is minimal (for all notions of minimality that we consider in this paper) and a subinstance of $A$,
	\item the union of the members of \multicore{A} is  equivalent to $A$,
	\item $A$ and $B$ are equivalent if and only if $\multicore{A}=\multicore{B}$, up to renaming of nulls, and
\item any instance which is  equivalent to $A$ and which is minimal in the sense of having no equivalent subinstance must be an image of the union of the members of  \multicore{A}. In particular, all such instances can be found, up to renaming of nulls, from (the union of) \multicore{A}.
\end{enumerate}
 %
 We also apply the analysis of na\"{i}ve evaluation of existential positive queries with Boolean universal guards from \cite{Gheerbrant:2014:NEQ:2691190.2691194} and show that such queries can be evaluated on the smaller members in \multicore{A} rather than on $A$ itself. Finally, we extend the analysis to OCWA* and show that, by resolving a question of  ``redundant annotation'', the function \multicore{A} can be extended also to annotated instances, yielding similar properties for OCWA*. 
In summary, the contributions of this paper are:
\begin{itemize}
\item A new semantics OCWA* which properly extends PCWA and OCWA with annotated instances.
\item Characterization and analysis of semantic implication and equivalence for PCWA, OCWA with annotated nulls, and OCWA*.
\item Negative results for the existence of unique minimal instances in PCWA
and OCWA*.
\item A new concept of `PCWA-core' for PCWA; and in terms of it, 
\item a new `powerset canonical representative function' \multicore{-} for PCWA and OCWA*, with the properties listed above.
%
%
%
\item An analysis of `annotation redundancy' in OCWA*.
\end{itemize}

The paper is organised as follows. 
In Section \ref{sec:prelims} we give preliminaries and introduce known semantics for incomplete DBs.
In Section \ref{sec:our-semantics} we present our OCWA* semantics and give its basic properties. 
In Section \ref{sec:EqandMinforOCWA} we study semantic implication and equivalence for OCWA*.
In Section \ref{sec: minimality-problems} we show the non-existence of a subinstance minimal representative function for PCWA and, consequently, for OCWA*. 
In Section \ref{sec:minforpcwa} we move to positive results for PCWA and then extend them
in Section \ref{sec:minforocwa} for the general case of OCWA*.




\section{Preliminaries} 
\label{sec:prelims}

We use boldface for lists and tuples; thus \x\ instead of $\vec x$ or $\overrightarrow{x}$. $\mathbb{N}^+$ is the set of positive (non-zero) natural numbers. $\mathcal{P}^+(A)$ is the set of non-empty subsets of $A$. $\mathcal{P}^{\mathrm{fin}}(A)$ is the set of finite subsets of $A$. If $S$ is a set of instances then $\overline{S}^{\cup}$ denotes the closure of $S$ under binary unions.

\subsection{Incomplete Databases}
We assume that we are working with a fixed database schema. 
Let $\mathsf{Const}$ and $\mathsf{Null}$ be countable sets of constants and labeled nulls. 
For the sake of readability, we will use lower case letters late in the alphabet for nulls instead of the more common $\bot$. Lower case letters $a$, $b$, $c$, $d$ will be used for constants. 
An (incomplete) \emph{instance} $A$ is a database instance whose (active) domain is a subset of $\mathsf{Const}\cup \mathsf{Null}$.
A \emph{complete instance} $I$ is an instance without nulls. (This is also known as a \emph{ground} instance.) We write $\mathcal{D}$ for the set of all  instances and $\mathcal{C}$ for the set of all complete  instances. We use upper case letters $A$, $B$, etc.\ from the beginning of the alphabet for instances in general, and upper case letters $I$, $J$, etc.\ from the middle of the alphabet for instances that are explicitly assumed to be complete.

Following \cite{DBLP:journals/jcss/LibkinS11} an \emph{annotated instance} is an instance
where each \emph{occurrence} of a constant or null is annotated with 
either $o$,  standing for \emph{open}, or $c$,  standing for \emph{closed}. 
The added expressivity is used to define more fine-grained semantics for incomplete databases. 
%

\subsection{Homomorphisms and Disjoint Unions}
\label{subsection: copowers}

We use the terms ``homomorphism'' and ``isomorphism'' to mean database homomorphism and database isomorphism, respectively, and we distinguish these from  ``structure'' homomorphisms. Explicitly,  if $A$ and $B$ are instances --- whether incomplete or complete, annotated or not --- a \emph{structure homomorphism} $h:A\rightarrow B$ is a function from the active domain of $A$ to the active domain of $B$ such that for every relation symbol $R$, if a tuple \BF{u} is in the relation $R$ in $A$ then the tuple $h(\BF{u})$ is in the relation $R$ in $B$. 
We write $\mathsf{Str}(A,B)$ for the set of structure homomorphisms from $A$ to $B$. 
A \emph{structure isomorphism} is an invertible structure homomorphism.

If $P\subseteq \mathsf{Const}\cup\mathsf{Null}$ and $h$ is a structure homomorphism we say that  $h$ \emph{fixes $P$ pointwise} if $h(p)=p$ for all $p\in P$ on which $h$ is defined. We say 
that $h$ \emph{fixes $P$ setwise} if it restricts to a bijection on the subset of $P$ on which it is defined. 
%
%

  A \emph{(database) homomorphism} from $A$ to $B$ is a structure homomorphism that fixes $\mathsf{Const}$ pointwise. We write \homset{A}{B} for the set of  homomorphisms from $A$ to $B$.
%
A \emph{(database) isomorphism} is an invertible  homomorphism.

A \emph{subinstance} of $A$ is an instance $B$ with an inclusion homomorphism $B\hookrightarrow A$---that is, with a homomorphism that fixes $\mathsf{Const}\cup\mathsf{Null}$ pointwise. $B$ is a  \emph{proper} subinstance if $A\neq B$.    We shall often be somewhat lax with the notion of a subinstance and regard $B$ as  a subinstance if it is so up to renaming of nulls, that is to say, up to (database) isomorphism. 
If we need to insist that the homomorphism $B\hookrightarrow A$ is an inclusion we say that $B$ is a \emph{strict} subinstance.   

If $h:A\rightarrow B$ is a structure homomorphism then the \emph{image} $h(A)$ of $h$ is the subinstance of $B$ defined by the condition that $\BF{v}$ is in the relation $R$ in $h(A)$ if there exists \BF{u} in $R$ in $A$ so that  $h(\BF{u})=\BF{v}$. If $h(A)=B$ we say that $h$ is  \emph{ strongly surjective}  and write $h:A\twoheadrightarrow B$.  If $h$ is not a structure isomorphism we say that $h(A)$ is a  \emph{proper} 	 image.    

A \emph{reflective subinstance} of $A$ is an instance $B$ with  an inclusion homomorphism $m:B\hookrightarrow A$ and a strongly surjective  homomorphism $q:A\twoheadrightarrow B$ such that $q\circ m$ is the identity  on $B$. Again, we often say that $B$ is a reflective subinstance if it is so up to renaming of nulls, and say that it is a strict reflective subinstance if we want to insist that $m$ is an inclusion, rather than just an injective homomorphism. 

If $H=\cterm{h_i:A\rightarrow B}{i\in S}$ is a family of  homomorphisms we say that $H$ is a \emph{covering family}, or simply a \emph{cover}, if $B=\bigcup_{i\in S} h_i(A)$.
We say that $A$ \emph{covers} $B$ if \homset{A}{B}
is a cover. 
If $H=\cterm{h_i:A_i\rightarrow B}{i\in S}$ is a family of homomorphisms with the same codomain we say that $H$ \emph{jointly covers} $B$ if $B=\bigcup_{i\in S} h_i(A_i)$ 


If $A$ is an incomplete instance, a \emph{freeze} of $A$ is, as usual, a complete instance $\overline{A}$ together with a structure isomorphism between $A$ and $\overline{A}$ that fixes the constants in $A$. Whenever we take a freeze of an instance, we tacitly assume that it is ``fresh'', in the sense that the new constants in it do not occur in any other instances currently under consideration (that is, usually, that have been introduced so far in the proof). 


We define the \emph{null-disjoint} union $A\sqcup_{\mathsf{Null}}B$ of two instances $A$ and $B$ to be the instance obtained by renaming whatever nulls necessary to make sure that $A$ and $B$  have no nulls in common, and then taking the union of the result. As such, the null-disjoint union is only defined up to isomorphism. The key property of the null-disjoint union is the 1-1 correspondence
$\homset{A\sqcup_{\mathsf{Null}}B}{C}\cong \homset{A}{C}\times \homset{B}{C}$
between homomorphisms from $A\sqcup_{\mathsf{Null}}B$ and pairs of homomorphisms from $A$ and $B$.

The definition extends to $n$-ary and infinitary null-disjoint unions. (Infinitary null-disjoint unions are, strictly speaking, not database instances in so far as they are not finite, but they are an occasionally useful technical extrapolation, and we trust that they will cause no confusion in the places where we make use of them.) We shall mostly be considering the null-disjoint union of an instance with itself. For $n\in \mathbb{N}^+\cup\{\infty\}$, we abuse notation and simply write $A^n$ for the null-disjoint union of $A$ with itself $n$ times, with the property that  $\homset{A^n}{C}\cong \prod_{i=1}^n\homset{A}{C}$.
We denote by $\pi_m:A\rightarrow A^n$, for $m\in \mathbb{N}^{+}$ smaller or equal to $n$, the  homomorphism that sends $A$ to the $m$th copy of it in $A^{n}$. 
If  $f:A^{n} \rightarrow C$ is a  homomorphism we write $f=\left\langle f_1,\ldots,f_n\right\rangle$ where $f_i=f\comp \pi_i:A\rightarrow C$.
We denote by $\nabla:A^{n}\rightarrow A$ the  homomorphism that corresponds to the $n$-tuple of identity homomorphisms $A\rightarrow A$. That is to say, $\nabla$  identifies all copies in $A^{n}$ of a null in $A$ with that null.

\subsection{Semantics of Incomplete Databases}
\label{sec:semanticprelims}

A \emph{semantics} is a function $\sem{-}:\mathcal{D}\rightarrow \mathcal{P}^{+}(\mathcal{C})$ which assigns a non-empty set \sem{A} of complete instances to every instance $A$. 
We say that $A$ \emph{represents} \Rep[]{A}.


A semantics \sem{-} induces a preordering on $\mathcal{D}$ by $A \leq B\Leftrightarrow \sem{A}\subseteq \sem{B}$\footnote{Note that this is the opposite of the standard order as defined  in e.g.\   \cite{Gheerbrant:2014:NEQ:2691190.2691194}}. We say that $A$ and $B$ are \emph{semantically equivalent}, and write $A\equiv B$,  if $A\leq B$ and $B\leq A$.  Accordingly,  $A\equiv B \Leftrightarrow \Rep[]{A}=\Rep[]{B}$.
The semantic equivalence class of an instance is denoted using square brackets: $[A]:=\cterm{B\in \mathcal{D}}{A\equiv B}$.

A \emph{representative function } (\emph{cf.}\ \emph{representative set, canonical function} in \cite{Gheerbrant:2014:NEQ:2691190.2691194}) is a function $\chi:\mathcal{D}\rightarrow \mathcal{D}$ 
which picks a representative of each semantic equivalence class. 
 We shall be content with $\chi(A)$ being defined up to  isomorphism. 
A representative function  $\chi$ is \emph{subinstance minimal} if $\chi(A)$ is a subinstance of all members of $[A]$.

Next, we briefly recall the established semantics OWA, CWA,
the Closed Powerset semantics of \cite{Gheerbrant:2014:NEQ:2691190.2691194},  and the Open and Closed World Assumption as defined by Libkin and Sirangelo \cite{DBLP:journals/jcss/LibkinS11}. 

\subsubsection{Open World Approach: OWA}

Under OWA (Open World Assumption)  an instance $A$ represents the set of complete instances to which $A$ has a (database)  homomorphism;
$\Rep[OWA]{A}=\cterm{I\in \mathcal{C}}{\homset{A}{I}\neq \emptyset }$.

Consequently, \Rep[OWA]{A} is closed under structure homomorphisms that fix the constants in $A$ pointwise, in the sense that if $I\in\Rep[OWA]{A}$ and $I\rightarrow J$ is a structure homomorphism that fixes the constants in $A$, then $J\in \Rep[OWA]{A}$.  
It is well known (see e.g.\ \cite{DBLP:journals/tods/FaginKP05}) that the function $\mathsf{Core}(-)$ that maps each instance to its core is a subinstance minimal representative function. 

\subsubsection{Closed World Approach: CWA}

Under CWA (Closed World Assumption) an instance $A$ represesents the set of its images; 
$\Rep[CWA]{A}=\cterm{I\in \mathcal{C}}{\textnormal{there exists } h:A\twoheadrightarrow I }$

Note that \Rep[CWA]{A}  is closed under strongly surjective structure homomorphisms that fix the constants in $A$ pointwise. 
%
%
Clearly, the only possible representative function (up to isomorphism, as usual) is the identity.

\subsubsection{Closed Powerset: PCWA}

Under \emph{Closed Powerset semantics} (PCWA) \cite{Gheerbrant:2014:NEQ:2691190.2691194},  $A$ represents the set of its CWA-interpretations closed under union; 
$\Rep[PCWA]{A}=\overline{\Rep[CWA]{A}}^{\cup}=\cterm{I_1\cup\ldots\cup I_n}{n\in \mathbb{N}^+,\ I_1,\ldots , I_n\in \Rep[CWA]{A}}$
Consequently, \Rep[PCWA]{A} is closed under unions and under strongly surjective homomorphisms that fix the constants in $A$ pointwise. 
Note that in \cite{Gheerbrant:2014:NEQ:2691190.2691194} this semantics is denoted $(|A|)_{\mathsf{CWA}}$
We recall the following from \cite[Thm 10.1]{Gheerbrant:2014:NEQ:2691190.2691194};
\begin{proposition}\label{Proposition: Libkin chara}
$A\leq_{\mathsf{PCWA}} B$ iff there exists a cover from $B$ to $A$.
\end{proposition}
Thus, $A\equiv_{\mathsf{PCWA}} B$ iff there exists a cover both from $B$ to $A$ and from $A$ to $B$.
The existence of minimal representative functions for PCWA is the subject of \ref{sec: minimality-problems} and  \ref{sec:minforpcwa}.

\begin{remark}The semantics GCWA introduced in 
\cite{Hernich:2010:ANQ:1804669.1804688}
defines  $\Rep[GCWA]{A}$ as the set of unions of \emph{minimal} images of $A$. In \cite{Gheerbrant:2014:NEQ:2691190.2691194} $\Rep[GCWA]{A}$ is denoted by $\left(|A|\right)^{\mathrm{min}}_{\mathsf{CWA}}$.
%
As with OWA,  $\mathsf{Core}(-)$   is a minimal representative function for GCWA (see \cite{Gheerbrant:2014:NEQ:2691190.2691194})

$\Rep[GCWA]{A}$ is not in general closed under strong surjections preserving the constants in $A$ (cf.\ \cite[9.1]{Gheerbrant:2014:NEQ:2691190.2691194}).  It therefore cannot be represented in the semantics introduced in \ref{sec:our-semantics} below.
\end{remark}

\subsubsection{Mixed Approach: OCWA$^{\textnormal{LS}}$}
\label{subsection: OCWA(LS)}

Let $A$ be an annotated instance, i.e.\ such that each occurrence of a constant or null is annotated as open or closed. 
Under OCWA$^{\textnormal{LS}}$  (Open and Closed World Assumption - Libkin/Sirangelo ) the set of complete instances represented by $A$   is defined in two steps as follows~\cite{DBLP:journals/jcss/LibkinS11}:
for all complete instances $I$,  $I\in \Rep[OCWA^{\textnormal{LS}}]{A}$ if 
\begin{itemize} 
\item[(i)] there exists a homomorphism $h:A \rightarrow I$; and 
\item[(ii)] for every $R(\mathbf{t})$ in $I$
	there exists a 
	$R(\mathbf{t}')$ in $A$ such that 
	 $h(\mathbf{t}')$ and $\mathbf{t}$ 
	agree on all positions annotated as closed in $\mathbf{t}'$. 
\end{itemize}
OCWA$^{LS}$ is subsumed by a more expressive semantics which we define next. 

%

\section{Our Semantics: OCWA*}
\label{sec:our-semantics}

In this section we propose the semantics OCWA*  for  annotated instances as a properly more expressive version of both OCWA$^{\textnormal{LS}}$and PCWA. The semantics OCWA* presupposes that instances are annotated according to certain conditions, which we define first:

\begin{definition}
We say that an annotated instance $A$ is  in \emph{normal form} if:
\begin{enumerate}
\item all occurrences of constants in $A$ are annotated as closed; and 
\item  all occurrences in $A$ of a null agree on the annotation of that null. 
\end{enumerate}  
\end{definition}

The following then allows us to restrict attention to instances in normal form without loss of generality with respect to OCWA$^{LS}$.

\begin{proposition}\label{proposition: Libkinrewrite}
Let $A$ be an annotated instance. Then there exists an annotated instance $A'$ in normal form such that $\Rep[OCWA^{LS}]{A}=\Rep[OCWA^{LS}]{A'}$.
\end{proposition}
\begin{proof}
For any atoms that contain open constants or open nulls annotated as closed elsewhere, change the annotation to `closed' and add a copy of the atom where those terms are replaced by fresh open nulls. 

\end{proof}

\begin{definition}If $A$ is a normal form annotated instance and $B$ is an instance, an \emph{RCN-cover} $H:A\rcncovers B$ is a set $H\subseteq \homset{A}{B}$ such that the homomorphisms in $H$  are jointly strongly surjective and agree on the closed nulls of $A$. 
\end{definition}

\begin{definition}[OCWA*]
Let $A$ be a  annotated instance in normal form. Then $A$ represents those complete instances under OCWA* that it RCN-covers;
$\Rep{A}=\cterm{I\in \mathcal{C}}{\exists H\!:\!A\rcncovers I}$.
\end{definition}

\begin{remark}
The definition of \Rep{A} could equivalently be given as the set 
of finite unions $h_1(A)\cup \ldots\cup h_n(A)$ of complete images of $A$ such that the homomorphisms $h_1,\ldots,h_n$ agree on the closed nulls of $A$.
Thus OCWA* lies within what \cite{Gheerbrant:2014:NEQ:2691190.2691194} call \emph{Powerset semantics}; that is, semantics that are defined in terms of a relation from instances to sets of complete instances (certain finite sets of valuations, in this case) and a relation from sets of  complete instances to complete instances (unions, in this case).
\end{remark}

OCWA* properly extends OCWA$^{\textnormal{LS}}$in the following sense:

\begin{theorem}\label{thm:rep_A-and-rep_C}
\begin{enumerate}
\item For every normal form annotated instance $A$ one can compute in time linear in $|A|$ 
a normal form annotated instance $A'$ such that $\Rep[OCWA^{LS}]{A} = \Rep[OCWA^{LS}]{A'} = \Rep{A'}$.
\item
There is a normal form annotated instance $A$ such that for every $A'$ it holds that
$\Rep{A} \neq \Rep[OCWA^{LS}]{A'}$.	
\end{enumerate}
\end{theorem}
\begin{proof}

 (1) Given $A$, extend it to  a new instance $A'$ by: for each atom $R(\mathbf{t})$ in $A$ add an atom $R(\mathbf{t}')$ where $\mathbf{t}'$ has every \emph{occurrence} of an open null in $\mathbf{t}$ replaced  by a fresh open null. It is then  straightforwardly verified that $\Rep[OCWA^{LS}]{A} = \Rep[OCWA^{LS}]{A'} = \Rep{A'}$. 

  (2) Consider the annotated instance $A = \{R(a^c, x^o, x^o )\}$.
   The instances in $\Rep{A}$ contain only tuples where the second and third coordinate are equal. However, the definition of OCWA$^{LS}$ requires only that one tuple in each instance from $\Rep[OCWA^{LS}]{A}$  respects this equality. Since there is no bound on the size of instances in $\Rep{A}$, there is no $A'$ such that $\Rep[OCWA^{LS}]{A'} = \Rep{A}$.
\end{proof}

Regarding PCWA, if $A$ is a normal form annotated instance without any closed nulls, then an RCN-cover $A\rcncovers C$ is simply a cover, since there are no closed nulls to agree upon. Thus PCWA is OCWA* restricted to instances without closed nulls. Explicitly, let $A$ be an un-annotated instance, and let its \emph{canonical annotation} be that which annotates each constant as closed and each null as open. Then we have:
\begin{proposition}
Let $A$ be an (un-annotated) instance and let $A[]$ be that instance with canonical annotation. Then $\Rep{A[]}=\Rep[PCWA]{A}$.
\end{proposition}


For the rest of this paper we assume that all annotated instances are in normal form. This allows us to introduce some notational conventions that simplify the study of RCN-covers on such instances. 
We also switch to annotating nulls by using lower and upper case instead of superscripts, 
since this allows us to more clearly emphasize the distinguished status of the closed nulls. 
%
We introduce the following conventions:

-- Open nulls are written in lower case,  $x$, $y$, $z$.
 Closed nulls are written in upper case,  $X$, $Y$, $Z$. 
(All instances are in normal form, so no null may occur both in lower and upper case in an instance.)

-- We display the closed nulls of an instance together with the instance; so that $A[\X]$ is an annotated instance where \X\ is a listing of the closed nulls of the instance. Thus \X\ can be the empty list. We allow ourselves to treat \X\ as the set of closed nulls of $A$ when convenient. It is a list for purposes of substitution. In particular:

-- If $\mathbf{t}$ is a list of constants or nulls,   $A[\mathbf{t}/\X]$ is the instance obtained by replacing \X\ with $\mathbf{t}$. If clear from context, we use  $A[\mathbf{t}]$ as shorthand. 

-- Let $n\in \mathbb{N}^+\cup\{\infty\}$. Recall from  \ref{subsection: copowers} that we, for an un-annotated instance $A$, write $A^n$ as a shorthand for the $n$-ary null-disjoint union of $A$ with itself. For an annotated instance  $A[\X]$  with closed nulls \X, we extend this notation and write $A^n[\X]$ for the $n$-ary \emph{open-null-disjoint} union; that is, the result of taking the union of $n$ copies of  $A[\X]$ where the \emph{open} nulls have been renamed so that no two copies have any open nulls in common. Accordingly,  a homomorphism $A^n[\X]\rightarrow C$ corresponds to an $n$-tuple of homomorphisms $A[\X]\rightarrow C$ that agree on the closed nulls \X.

We close this section by displaying some equivalent definitions of \Rep{A[\X]}, including in terms of CWA and PCWA, which will be made use of in the sequel.
Note that for $n\in \mathbb{N}^+\cup \{\infty\}$, the family 
$\cterm{\pi_m:A[\X]\rightarrow  A^n[\X]}{m\leq n,\ m\in \mathbb{N}^+} $
forms a RCN cover from $A[\X]$ to $A^n[\X]$.


\begin{theorem}\label{thm: OCWA rep defs}
Let $A[\X]$ be an annotated instance and $I$ a complete instance. 
The following are equivalent:
\begin{enumerate}
\item $I\in \Rep{A[\X]}$, i.e\ there exist an RCN-cover $A[\X]\rcncovers I$; 
\item $I\in \bigcup_{n\in \mathbb{N}^+}\Rep[CWA]{A^{n}[\X]}$; 
\item $I\in \Rep[CWA]{A^{\infty}[\X]}$; and
\item $I  \in \bigcup_{\BF{d}\in \mathsf{Const}^{k}}\Rep[PCWA]{A[\BF{d}/\X]}$
where $k$ is the length of \X.
\end{enumerate}
\end{theorem}

\begin{corollary}
$\Rep{A[\X]}$ is closed under strongly surjective structure homomorphisms that fix the constants in $A[\X]$ pointwise. 
\end{corollary}

We now proceed to the study of implication and equivalence OCWA*.

\section{OCWA*: Implication, Equivalence}
\label{sec:EqandMinforOCWA}

%
%
  %


Since RCN-covers are closed under left composition with strong surjections, we have (by \ref{thm: OCWA rep defs}) that $\Rep[]{A[\X]}  \subseteq \Rep[]{B[\Y]} $ iff there is an RCN-cover from $B[\Y]$ to $ A^n[\X]$, for all $n\in \mathbb{N}^+$, or, equivalently, that there is an RCN-cover from  $B[\Y]$ to $A^{\infty}[\X]$ . We display this and  show that $n$ can be bounded by a number depending on $B$, or indeed that $n$ can be bounded by $2$ if one considers RCN-covers of a particular form. Note that the following theorem can also be applied to OCWA$^{LS}$ via the translations of \ref{proposition: Libkinrewrite} and \ref{thm:rep_A-and-rep_C}.
\begin{theorem}\label{thm:cover-chara}
Let $A[\X]$ and $B[\Y]$ be annotated instances. The following are equivalent:
%
\begin{compactenum}[\it (i)]
\item\label{thm:cover-chara1} $\Rep{A[\X]}  \subseteq \Rep{B[\Y]}$.
\item\label{thm:cover-chara2} There is an RCN-cover from  $B[\Y]$ to $ A^n[\X]$, for all  $n\in \mathbb{N}^+$.
\item\label{thm:cover-chara5} There is an RCN-cover  from $B[\Y]$ to $A^{\infty} [\X]$
\item\label{thm:cover-chara6} There is a strongly surjective  homomorphism   from $B^{\infty} [\Y]$ to $ A^{\infty} [\X]$.
\item\label{thm:cover-chara3} There is an RCN-cover from $B[\Y]$ to $ A^{n+1}[\X] $ where $n$ is the number of closed nulls in $B[\Y]$, i.e.\ the length of \Y.
\item\label{thm:cover-chara4}  There exists a RCN-cover $H$ from $B[\Y]$ to $A^2[\X]$ such that $H$ contains at least one  homomorphism  $h$ which factors through $\pi_1: A[\X]\rightarrow A^2[\X]$. 
\end{compactenum}
 
\end{theorem}
\begin{proof}
%
%
\ref{thm:cover-chara3}$\Rightarrow$\ref{thm:cover-chara4} : Let $n$ be the length of \Y, and let $H$ be  an RCN-cover  from $B[\Y]$ to $A^{n+1} [\X]$.
Choose an $h$ in $H$. There are more copies of $A[\X]$ in $A^{n+1}[\X]$ than there are \Y{}s, so we can assume that for all closed nulls $Y_i$ in $B[\Y]$, if $h(Y_i)$ is in the $n+1$th   copy, then $h(Y_i)$ is either a closed null or a constant. Then the composite
$h'=\left\langle \pi_1, \ldots, \pi_n,\pi_n \right\rangle\comp h :B[\Y]\rightarrow A^{n+1}[\X]\rightarrow A^{n+1}[\X]$
agrees with $H$ on all \Y, so 
$H'=H\cup\{h'\}$
is an RCN-cover. 
Now, if we compose $H'$ with the strong surjection
$\left\langle \pi_1, \ldots, \pi_1,\pi_2 \right\rangle : A^{n+1}[\X]\rightarrow A^2[\X]$
which sends the $n$ first copies of $A^{n+1}[\X]$ to the first copy in $ A^2[\X]$ and the $n+1$th copy of $A^{n+1}[\X]$ to the second in $A^{2}[\X]$, we obtain an RCN-cover of $A[\X]\rightarrow A^2[\X]$ in which the map $\left\langle \pi_1, \ldots, \pi_1,\pi_2 \right\rangle\comp h'$ factors through $\pi_1: A[\X]\rightarrow A^2[\X]$.

\ref{thm:cover-chara4}$\Rightarrow$\ref{thm:cover-chara2}: Let $n$ be given, and let $H$ be an RCN-cover  from $B[\Y]$ to $A^2[\X]$ such that $h\in H$ factors through $\pi_1: A[\X]\rightarrow A^2[\X]$. 
%
%
For $1\leq i \leq n$, $\pi_1:A[\X]\rightarrow  A^n[\X]$ and $\pi_i:A\rightarrow  A^n[\X]$ is a pair of homomorphisms that agree on closed nulls, so correspond to a homomorphism $\left\langle \pi_1, \pi_i \right\rangle:A^2[\X]\rightarrow  A^n[\X]$.  The family \cterm{\left\langle \pi_1, \pi_i \right\rangle}{1\leq i \leq n} of such homomorphisms is an RCN-cover from $A^2[\X]$ to $  A^n[\X]$. 
The composite of this cover  with $H$ is RCN, since for any closed null $Y_i$ in $B[\Y]$, $h'\in H$,  $1\leq i \leq n$, we have that $\left\langle \pi_1, \pi_i \right\rangle (h'(Y_i))= \left\langle \pi_1, \pi_i \right\rangle (h(Y_i))=\pi_1(h(Y_i))$.

The remaining implications are straightforward.
\end{proof}

From \ref{thm:cover-chara} we can derive two guess-and-check algorithms to decide containment between annotated instances. On the one hand, we may construct $ A^{n+1}[\X]$, where $n$ is the length of \Y, guess a set of homomorphisms from $B[\Y]$ to this instance, and check that it is an RCN-cover. Alternatively, we may avoid this blowup of $A[\X]$ by constructing $A^2[\X]$, guessing a homomorphism $h$ from $B[\Y]$ to $A[\X]$ as well as a set of homomorphisms $H$ from $B[\Y]$ to $A^2[\X]$, and checking that $\{h\} \cup H$ is an RCN-cover.

\paragraph*{Complexity analysis}  Since the instance $A^{n+1}[\X]$ has size at most $|A[\X]| \times (|\Y|+1)$, and the number of homomorphisms in any non-redundant cover is bounded by the number of tuples in the target instance, the complexity of this problem stays in \NP. For \NP-hardness, we adapt the reduction of 3-colourability for graphs to the problem of deciding whether a given graph has a homomorphism into $K_3$, the complete graph on three vertices. It is easy to see that any homomorphism from a graph with at least one edge into $K_3$ extends to a cover of $K_3$. Therefore, the problem of deciding 
$\leq_{\mathsf{PCWA}}$, and consequently $\leq_{\mathsf{OCWA*}}$, is likewise \NP-complete. 
It follows that the problem of deciding, given two instances $A$ and $B$, whether $A$ is a minimal equivalent instance for $B$ given a partial order among instances, belongs to the class \DP, as it involves checking the non-existence of a smaller instance. In other words, deciding semantic implication and equivalence for annotated instances has the same complexity as the homomorphism problem.

%

\section{Issues with Minimality in OCWA* }
\label{sec: minimality-problems}

In this section and the next  we study the notion of OCWA* semantic equivalence and the question of whether, or to what extent, there exists a unique ``best'' annotated instance to choose among those that are semantically equivalent.    
For motivation and illustration, we first recall the situation in  OWA in some more detail. 
%
%
  It is well known
  that $A\equiv_{\mathsf{OWA}}B$ if and only if $A$ and $B$ are ``homomorphically equivalent'', that is, if there exists a homomorphism both from $A$ to $B$ and from $B$ to $A$. Furthermore,  there is, up to isomorphism, a least subinstance of $A$ to which it is homomorphically equivalent, known as the \emph{core} of $A$. 
Instances $A$ and $B$ are homomorphically equivalent if and only if their cores are isomorphic.
Moreover, as a consequence of  being the least homomorphically equivalent subinstance of $A$, the core of $A$ is also the least reflective subinstance of $A$, and the least homomorphically equivalent image of $A$. Thus there are three quite natural notions of minimality to which the core is the answer in OWA.
We say that an instance is \emph{a core} if it is its own core, i.e.\ if it has no homomorphically equivalent subinstances. Cores can be characterized as those instances $C$ with the property that any homomorphism $C\rightarrow C$ must be an isomorphism.  (See \cite{HELL1992117,foniok-thesis, DBLP:journals/tods/FaginKP05} for more about cores.)

We show now  that for OCWA* there does not in general exist least semantically equivalent instances in any of the three senses above. We then turn to the question of whether a `good' representative function can nevertheless be found, first for PCWA and then for OCWA* in general.
We begin by fixing some terminology.
\begin{definition}\label{def:minimality terms}
Let $A$ and $B$ be instances. In the context of a given semantics, we say that:
\begin{enumerate}
\item  $B$ is  \emph{sub-minimal} (subinstance minimal) if there are no proper   semantically equivalent  subinstances of  $B$;  
\item  $B$ is  \emph{rfl-minimal} (reflective subinstance minimal) if there are no proper   semantically equivalent  reflective subinstances of  $B$;  
\item $B$ is a \emph{least} semantically equivalent (reflective) subinstance of $A$ if $B\equiv A$ and  $B$ is a (reflective) subinstance of all semantically equivalent (reflective) subinstances of $A$;
\item $B$ is   \emph{ img-minimal} (image minimal) if there are no proper semantically equivalent images of $B$, and finally;  
 \item $B$ is a  \emph{least} semantically equivalent image of $A$ if $B\equiv A$ and for all semantically equivalent images $C$ of $A$, $B$ is an image of $C$.  
\end{enumerate}
\end{definition}

%
We show by the examples that follow that in PCWA, and hence in  OCWA*,  least semantically equivalent subinstances, reflective subinstances, and images do not in general exist, and that when they do, they need not coincide. In the examples all instances consist of  nulls only. 

\begin{example}\label{Example: no least sub}
$B_1$ and $C_1$ are non-isomorphic PCWA-equivalent reflective subinstances of $A_1$. Both   $B_1$ and $C_1$ are sub-minimal and rfl-minimal.
\begin{center}
\scalebox{0.8}{\begin{tabular}{ l | c c r }
 $A_{1}$&R&&\\
\hline
   &x & x & y \\
   &x & x & x  \\
   &v & w & w  \\
   &v&v&v\\
   &z&z&r\\
   &z&s&s\\
   &z&z&z\\
\end{tabular}}
\quad
\scalebox{0.8}{\begin{tabular}{ l | c c r }
 $B_{1}$&R&&\\
\hline
   &x & x & y \\
   &x & x & x  \\
   &v & w & w  \\
   &v&v&v\\
\end{tabular}}
\quad
\scalebox{0.8}{\begin{tabular}{ l | c c r }
 $C_{1}$&R&&\\
\hline
   &z&z&r\\
   &z&s&s\\
   &z&z&z\\
\end{tabular}}
\end{center}
\end{example}

\begin{example}\label{example:No least R-quotient}
The instances $B_2$ and $C_2$ are non-isomorphic PCWA-equivalent images of the instance  $A_2$. Both $B_2$ and $C_2$ are img-minimal. 
\begin{center}
\scalebox{0.8}{\begin{tabular}{ l | c c c c r }
$A_2$ &R&&&&\\
\hline
   &x & x & u & y & z \\
   &x & x &  x & x & z\\
   &x & x & x  & y & x\\
  &x & x & x  & x & x\\
   &v & p & p & r & s \\
   & p& p &  p & p & s\\
   &p & p & p  & r & p\\
  &p & p & p  & p & p\\
\end{tabular}}
\quad
\scalebox{0.8}{\begin{tabular}{ l | c c c c r }
$B_2$ &R&&&&\\
\hline
   &p & p & u & r & z \\
   &p & p &  p & p & z\\
   &v & p & p & r & s \\
   & p& p &  p & p & s\\
   &p & p & p  & r & p\\
  &p & p & p  & p & p\\
\end{tabular}}
\quad
\scalebox{0.8}{\begin{tabular}{ l | c c c c r }
$C_2$ &R&&&&\\
\hline
   &p & p & u & y & s \\
   &p & p & p  & y & p\\
   &v & p & p & r & s \\
   & p& p &  p & p & s\\
   &p & p & p  & r & p\\
  &p & p & p  & p & p\\
\end{tabular}}
\end{center}
\end{example}

\begin{example}\label{ex: different cores are different}The instance  $A_3$ has a least PCWA-equivalent reflective subinstance, a least PCWA-equivalent subinstance, and a least PCWA-equivalent image, consisting of the non-isomorphic instances $A_3$, $B_3$, and $C_3$,  respectively:
\begin{center}
\scalebox{0.8}{\begin{tabular}{ l | c c c c r }
$A_3$ & R&&&&\\
\hline
  &x & x' & y & y & z \\
  &v' & v & s  & t  & s\\
  &x & v &  u & u & u\\
  &x & x & x  & x & x\\
 &v & v & v  & v & v\\
\end{tabular}}
\quad
\scalebox{0.8}{\begin{tabular}{ l | c c c c r }
 $B_3$&R&&&&\\
\hline
  &x & x' & y & y & z \\
  &v' & v & s  & t  & s\\
  &x & x & x  & x & x\\
 &v & v & v  & v & v\\
\end{tabular}}
\quad
\scalebox{0.8}{\begin{tabular}{ l | c c c c r }
 $C_3$&R&&&&\\
\hline
   &w & x' & y & y & z \\
   &v' & w & s  & t  & s\\
   &w & w & w  & w & w\\

\end{tabular}}
\end{center}
\end{example}

We summarize:

\begin{theorem}\label{thm:negative results} In PCWA (OCWA*),
\begin{enumerate}
\item there exists an (annotated) instance $A$ for which there exists two non-isomorphic semantically equivalent sub-minimal subinstances; 
\item there exists an (annotated) instance $A$ for which there exists two non-isomorphic semantically equivalent img-minimal images; and 
\item there exists an (annotated) instance $A$ for which there exists two non-isomorphic semantically equivalent rfl-minimal reflective subinstances.  
\end{enumerate}
\end{theorem}

%
%

\section{Minimality in PCWA}
\label{sec:minforpcwa}

Recall from \ref{sec:semanticprelims} that a representative function  for a given semantics is a function $\chi:\mathcal{D}\rightarrow\mathcal{D}$ which chooses a representative for each equivalence class. That is to say, $A\equiv B\Leftrightarrow \chi(A)= \chi(B)$, for all  $A, B\in \mathcal{D}$, and
$\chi(A)\equiv A$, for all $A\in \mathcal{D}$.
Again, we only require that $\chi(A)$  is defined up to isomorphism, i.e.\ up to renaming of nulls.
Recall further that a representative function is subinstance minimal if $\chi(A)$ is a subinstance of  $A$ (up to isomorphism) for all $A\in \mathcal{D}$. Similarly, we say that a representative function is \emph{image minimal} if $\chi(A)$  is an image of $A$, and that it is \emph{reflective subinstance minimal} if $\chi(A)$  is a reflective subinstance of $A$. The canonical example is the $\mathsf{Core}$ function, which is a minimal representative function for OWA in all of these three senses. 

 \ref{thm:negative results} showed that there can be no minimal representative function for PCWA, for any of these three senses of ``minimal''.
However, we show that there is a function
$\multicore{-}:\mathcal{D}\rightarrow\mathcal{P}^{\mathrm{fin}}(\mathcal{D})$
that assigns a finite set $\{E_1, \ldots,E_n\}$ to each instance $A$  that is representative in the sense that $A\equiv_{\mathsf{PCWA}} B\Leftrightarrow \multicore{A}= \multicore{B}$, for all  $A, B\in \mathcal{D}$, and $ \displaystyle\bigcup_{E\in\multicore{A}}E\equiv_{\mathsf{PCWA}} A$, for all $A\in \mathcal{A}$;
and `minimal' in the sense that
\begin{itemize}
\item[--] $E$ is a reflective subinstance of $A$, for all $A\in\mathcal{D}$ and all $E\in \multicore{A}$, and 
\item[--] $E$ is semantically minimal in the strong sense that if $C\equiv_{\mathsf{PCWA}} E$ then $E$ is a reflective subinstance of $C$, for all $E\in \multicore{A}$. 
\end{itemize}
Thus the members of \multicore{A} are both sub-, img-, and rfl-minimal, in the sense of \ref{def:minimality terms}.
Furthermore, if $\multicore{A}=\{E_1, \ldots,E_n\}$ then
\begin{equation}
\Rep[PCWA]{A}=\overline{\Rep[CWA]{E_1}\cup\ldots\cup\Rep[CWA]{E_n}}^{\cup}.
\end{equation}

We propose \multicore{-}  as a form of  ``power core'' or ``multi-core'' function for PCWA; giving for each $A$ a finite set of PCWA-minimal instances which jointly embody the PCWA-relevant structure of $A$, analogously to the role that the single instance $\Core (A)$ plays in OWA. 
In addition to the properties just listed, we show the following as an instance of the usefulness of \multicore{-}.  
%
For any given instance $A$, the set of sub-minimal subinstances of $A$ is of course finite. But this set may have no overlap with  the set of sub-minimal subinstances of $B$, even if $A$ and $B$ are semantically equivalent. Thus it is, on the face of it, not obvious that the set $\mathrm{Min}([A]_{\mathsf{PCWA}})$ of sub-minimal members of the whole equivalence class $[A]_{\mathsf{PCWA}}$ must be finite (up to renaming of nulls). However, we show that any sub-minimal member  of $[A]_{\mathsf{PCWA}}$ must be an image of $ \bigcup_{E\in\multicore{A}}E$, establishing thereby that \multicore{A} both yields a finite bound on the size of $\mathrm{Min}([A]_{\mathsf{PCWA}})$, and a way to compute it.  

Moreover, we show in  \ref{sec:queries} that
for the class of queries known as \emph{existential positive with Boolean universal guards}, the so-called \emph{certain answers} can in fact be computed  directly from the elements in \multicore{A}, rather than from the larger $A$.
%

In the rest of \ref{sec:minforpcwa} we fix the semantics to be PCWA, and thus leave the subscripts implicit.

 \subsection{PCWA-cores}
%
Recall that
$A$ is a core if and only if every homomorphism $A\rightarrow A$ is an isomorphism. In analogy, we introduce the notion of PCWA-core as follows. 
\begin{definition}\label{def:ecore}
We say that an instance $A$ is a  \emph{PCWA-core} if every self-cover $H\subseteq \homset{A}{A}$ contains an isomorphism.   
\end{definition}
%
%
\begin{example}
$D=\{R(z,z,r), R(z,z,z)\}$ is  a PCWA-core, as the only endomorphism hitting $R(z,z,r)$ is the identity.  
The core of $D$ is $\{R(z,z,z)\}$. 
\end{example}
Accordingly, every core is a PCWA-core. It is also evident that cores have the property that if $C$ is a core and $A$ is any instance, then $A$ and $C$ are OWA semantically equivalent if and only if $C$ is a reflective subinstance of $A$. For PCWA-cores we have the following:
\begin{proposition}\label{lemma: E-cores are least}
Let $A\equiv B$ and assume that $A$ is a PCWA-core. Then $A$ is a reflective subinstance of $B$.
\end{proposition}
\begin{proof}
$\homset{B}{A}\comp\homset{A}{B}$ is a cover so it contains an isomorphism.
\end{proof}
Consequently, if two PCWA-cores are semantically equivalent, they are isomorphic.

\ref{sec: minimality-problems} introduced  three different notions of minimality with respect to semantic equivalence. We  relate these to each other and to the property of being a PCWA-core. 
\begin{proposition}\label{lemma: core imps}
Let  $A$ be an instance. The following implications hold and are strict. 
\begin{enumerate}
\item If $A$ is a PCWA-core then $A$ is sub-minimal and img-minimal.
\item If $A$ is sub-minimal or img-minimal then it is rfl-minimal. 
\end{enumerate}
%
\end{proposition}
\begin{proof}
1) follows from \ref{lemma: E-cores are least} and 2) is immediate.That the implications are strict is shown in Examples \ref{Example: no least sub} and \ref{ex: different cores are different}. Specifically, $C_1$ of \ref{Example: no least sub} is both sub-minimal and img-minimal, but it is not a PCWA-core. And $A_3$ of \ref{ex: different cores are different} is rfl-minimal, but neither sub- nor img-minimal. 
\end{proof}

In what follows it is convenient to fix a more compact notation for  atoms $R(\mathbf{t})$ that occur in an  instance $A$.
We primarily use the variable $k$ for atoms, and  write $k\mathbin{:}A$ for ``$k$ is an atom of $A$''. 
%
If \homo{f}{A}{B} is a homomorphism and $k=R(\mathbf{t})\mathbin{:}A$ then $f(k)=R(f(\mathbf{t}))$.

%

We recall  the notion of ``core with respect to a tuple'':
\begin{definition}\label{def:structure core}
Let $k\mathbin{:}A$. The \emph{core of $A$ with respect to $k$}, denoted $C^A_k$, is the least strict reflective subinstance of $A$ containing $k$. 
\end{definition}
The instance $C^A_k$ can be regarded as the ``core of $A$ with $k$ frozen'', and thus is unique, up to isomorphism. As a reflective subinstance, it comes with an injective homomorphism to $A$ and a strong surjection from $A$, which we write \homo{m_k}{C^A_k}{A} and \homo{q_k}{A}{C^A_k}, respectively.  When the instance $A$ is clear from context, we leave the superscript implicit and just write $C_k$.
%
%
%
%
%
%
%
 %
%
%
%
%
%
%
%
We display the following  for emphasis.
\begin{lemma}
Any homomorphism $h:C^A_k\rightarrow C^A_k$ that fixes $k$ must be an isomorphism. 
\end{lemma}
\begin{definition}\label{def:maximal atom}
We say that two atoms $k,k'\mathbin{:}A$ are \emph{endomorphism-equivalent}, and write $k\sim_A k'$, if there exist $f,g\in \homset{A}{A}$ such that $f(k)=k'$ and $g(k')=k$. 
We say that  $k\mathbin{:} A$ is \emph{(endomorphism-)maximal} if ``only equivalent atoms map to it''. That is, for all $k'\mathbin{:}A$ and $f\in \homset{A}{A}$, $f(k')=k$ implies that  $k\sim_A k'$. If $k$ is maximal we write \Max{A}{k}.
\end{definition}
%
%
\begin{lemma}\label{lemma:eq ks yield eq cores}
Let $A$ be an instance, and $k,k' \mathbin{:} A$. If $k\sim_A k'$ then $C_k\cong C_{k'}$.
\end{lemma}
\begin{proof}Suppose $f,g\in \homset{A}{A}$ such that $f(k)=k'$ and $g(k')=k$ and consider
 the diagram 
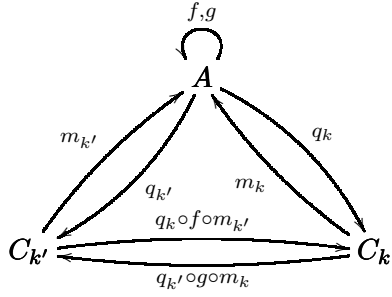
\[\bfig
\Atriangle|baa|/{@{->}@/^9pt/}`{@{->}@/^9pt/}`{@{>}@/^5pt/}/<600,600>[A`C_{k'}`C_{k};q_{k'}`q_k`q_k\comp f\comp m_{k'}]
\Atriangle|abb|/{@{<-}@/_5pt/}`{@{<-}@/_5pt/}`{@{<-}@/_5pt/}/<600,600>[A`C_{k'}`C_{k};m_{k'}`m_k`q_{k'}\comp g\comp  m_k]
\Loop(600,600){A}(ur,ul)_{f,g}
\efig\]
The homomorphism $h:=(q\comp f\comp m')\comp (q'\comp g\comp  m:C_k\rightarrow C_k)$ fixes $k$. So $h$ must be an isomorphism. 
By symmetry, we obtain that $C_k\cong C_{k'}$. 
\end{proof}

\begin{lemma}
If $k\mathbin{:}A$ is maximal, then $C_k$ is  a PCWA-core.
\end{lemma}
\begin{proof}
First note that for any instance $B$ and any  set of homomorphisms $H\subseteq \homset{B}{B}$, if $\overline{H}$ is the closure of $H$ under composition, then: 
1) $H$ is a cover if and only if $\overline{H}$ is a cover; and 
2) $H$ contains an isomorphism if and only if  $\overline{H}$ contains an isomorphism. 
%
Let $H\subseteq\homset{C_k}{C_k}$ be a cover, and assume without loss of generality that it is closed under composition. Then we can find $k':C_k$ and  $f\in H$ such that $f(k')=k$. Since $k$ is maximal in $A$ it is maximal in $C_k$, so there is a homomorphism $h:C_k\rightarrow C_k$ such that $h(k)=k'$. But then $f\comp h$ is an isomorphism, so $f$ must be an isomorphism as well.  
\end{proof}

Thus the maximal atoms of an instance determine a set of reflective subinstances which are PCWA-cores. We show that these are invariant under semantic equivalence. 
%
 %
\begin{theorem}\label{thm:sameatoms}
Let $H\subseteq \homset{A}{B}$ and $G\subseteq \homset{B}{A}$ be covers.  Let $k_B\mathbin{:}B$ be maximal. Then there exist $h\in H$ and $k_A\mathbin{:}A$ such that $k_A$ is maximal and $h(k_A)=k_B$. Moreover, the homomorphism $q_{k_B}\comp h\comp m_{k_A}: C^A_{k_A}\rightarrow C^B_{k_B}$ is an isomorphism.
\end{theorem}
\begin{proof}
First, we show that, more generally, whenever $A\equiv B$, it is the case that  for all $f:A\rightarrow B$ and all $k\mathbin{:}A$, if \Max{B}{f(k)} then \Max{A}{k}. 

For  suppose $g:A\rightarrow A$ and $k'\mathbin{:}A$ is such that $g(k')=k$. Choose $f':B\rightarrow A$ and $k''\mathbin{:}B$ such that $f'(k'')=k'$. Then $f\comp g\comp f'(k'')=f(k)$ so there is $f'':B\rightarrow B$ such that $f''(f(k))=k''$, whence $g\comp f'\comp f''\comp f(k)=k'$.  This establishes the first claim of the theorem. 

Next, let $\Max{A}{k_A}$, $\Max{B}{k_B}$, and $h\in H$ such that $h(k_A)=k_B$. 
Chose $k'\mathbin{:}B$ and $g\in G$ such that $g(k')=k_A$. Then $f\comp g(k')=k_B$, so there exists $f:B\rightarrow B$ such that $f(k_B)=k'$.
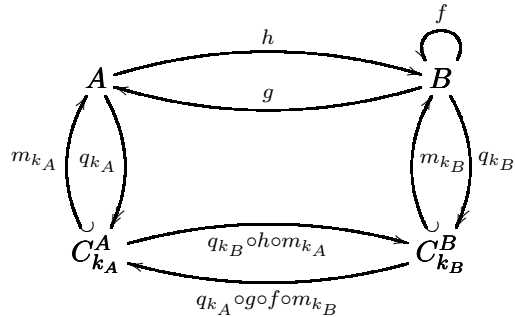
\[\bfig
\square/{@{>}@/^1em/}`{@{->>}@/^1em/}`{@{->>}@/^1em/}`{@{>}@/^1em/}/<1200,600>[A`B`C^A_{k_A}`C^B_{k_B};h`q_{k_A}`q_{k_B}`q_{k_B}\comp h\comp m_{k_A}]
\square/{@{<-}@/_1em/}`{@{<-^{)}}@/_1em/}`{@{<-^{)}}@/_1em/}`{@{<-}@/_1em/}/<1200,600>[A`B`C^A_{k_A}`C^B_{k_B};g`m_{k_A}`m_{k_B}`q_{k_A}\comp g\comp f\comp  m_{k_B}]
\Loop(1200,600){B}(ur,ul)_f
\efig\]
Then $q_{k_B}\comp h\comp m_{k_A}(k_A)=k_B$ and $q_{k_A}\comp g \comp f\comp m_{k_B}(k_B)=k_A$, whence their composites are isomorphisms. So they must themselves be isomorphisms. 
\end{proof}
%
%
%
%
%
%
%
%
%
Finally, we note the following property of PCWA-cores which will be used in the next section. 
\begin{lemma}\label{lemma:atomsbyelements}
An instance $A$ is a PCWA-core if and only if there exists $k\mathbin{:}A$ with the property  that for all $f\in \homset{A}{A}$, if ${k}$ is in the image of $f$ then $f$ is an isomorphism. 
\end{lemma}
\begin{proof}
Suppose $A$ is a PCWA core. For each maximal $k$, let $f_k:A\rightarrow A$ be the composition of $q_k:A\rightarrow C_k$ and $m_k:C_k\rightarrow A$. Then $\homset{A}{A}\comp \cterm{f_k}{\Max{A}{k}}$ is covering, so one of its homomorphisms, and hence one of the $f_k$s, must be an isomorphism. The converse is immediate.  
\end{proof}

\subsection{PCWA Multicores}

Consider the family \cterm{C_k}{\Max{A}{k}} of (strict) reflective subinstances of $A$. 
From the definition of maximality we have that  for any atom  $t:A$ there exists a maximal atom $k:A$ and an endomorphism $h:A\rightarrow A$ such that $f(k)=t$. Thus the family \cterm{C_k}{\Max{A}{k}} jointly covers $A$. 
%
%
%
%
Clearly, if we successively  remove any member of \cterm{C_k}{\Max{A}{k}}  that is a reflective subinstance of another  member, we will retain a subset that still jointly covers $A$. Thus we can summarize what we have so far with  the following.
\begin{theorem}\label{thm:multicore}\begin{enumerate}
\item For each $A\in \mathcal{D}$ there exists a finite set $\multicore{A}\subseteq \mathcal{D}$ such that:
\begin{enumerate}
\item for all $E\in \multicore{A}$, $E\cong C^A_k$ for some maximal $k:A$;
\item for all maximal $k:A$, there exists $E\in\multicore{A}$ such that $C^A_k$ is a reflective subinstance of $E$; and
\item for all $E,E'\in \multicore{A}$  if $E$ is a reflective subinstance of $E'$ then $E=E'$.
\end{enumerate} 
\item for given $A\in \mathcal{D}$ the set $\multicore{A}$ is unique with properties 1.(a)--1.(c), up to isomorphisms of its members. That is to say, if $X$ is another set satisfying properties 1.(a)--1.(c), then there exists a bijection ${f}:{\multicore{A}}\rightarrow{X}$ such that $f(E)\cong E$.
\item  $A\equiv B$ if and only if $\multicore{A}=\multicore{B}$, up to isomorphism of the members.
\end{enumerate} 
\end{theorem}
%

We refer to \multicore{A} as the \emph{multicore} of $A$. The multicore of an instance $A$ is only defined up to isomorphisms of its members, so we can assume without loss whenever it is convenient that no nulls are shared between those members; i.e.\ that for all $E,E'\in \multicore{A}$ we have $\mathrm{dom}(E)\cap \mathrm{dom}(E')\subseteq \mathsf{Cons}$. We also regard multicores as equal when their members are isomorphic.

Before proceeding, we characterize when a set of instances is (up to isomorphism) \multicore{A} for some $A$. 
We need the following lemma.
\begin{lemma}\label{lemma:whencoresarecovered}
Let $A$ be an instance and $B$ a  reflective subinstance of $A$. Let  $k:A$ and suppose that $k$ is maximal. Then $C_k$ is a reflective subinstance of $B$ if and only if there exists a homomorphism $f:B\rightarrow C_k$ and $k':B$ such that $f(k')=k$.  
\end{lemma}
\begin{proof}
The left-to-right is immediate.  Assume that there exists a homomorphism $f:B\rightarrow C_k$ and $k':B$ such that $f(k')=k$. Since $k$ is maximal in $A$ there exists a  homomorphism $g:C_k\rightarrow B$ such that $g(k)=k'$. But then $f\comp g$ fixes $k$, so it is an isomorphism. 
\end{proof}
%
%
\begin{theorem}\label{thm: the structure of multicores}
Let $\mathfrak{F}= \{C_1,\cdots,C_n\}$ be a family of  instances (with no nulls in common). The following are equivalent:
\begin{enumerate}
\item There exists an instance $A$ such that $\mathfrak{F}=\multicore{A}$ (up to isomorphism of the members).
\item
\begin{enumerate}
\item $C_i$ and $C_j$ have the same core (up to isomorphism) for all $i, j \leq n$, and
\item   there exists a selection of atoms $k_i\mathbin{:} C_i$, $1\leq i \leq n$, satisfying the condition that
if there exists \homo{h}{C_j}{C_i} such that $k_i$ is in the image of $h$, then $i=j$ and $h$ is an isomorphism.   
\end{enumerate}
\end{enumerate} 
\end{theorem}
\begin{proof}
  Assume $\mathfrak{F}=\multicore{A}$. Then we can regard \multicore{A} as \cterm{C_k}{k\in I} for a set $I$ of maximal $k:A$. 
Firstly, the core of $A$ is the core of $C_k$ for all $k\in I$.
Secondly, by \ref{lemma:whencoresarecovered}, if $\homo{h}{C_j}{C_k}$ such that $k$ is in the image of $h$ then $C_k$ is a reflective subinstance of $C_j$, whence by the definition of \multicore{A} we have that  $j=k$ and $h$ is an isomorphism. 
%

Assume  conditions in a) and b) are satisfied. b) ensures, together with \ref{lemma:atomsbyelements}, that $C_i$ is a PCWA core for all $i\in I$.
Let $A:=\displaystyle\bigcup_{k\in I} C_k$ (relying on the assumption that the members of $\mathfrak{F}$ have no nulls in common).  Since the $C_i$s share the same core, $C_i$ is a reflective subinstance of $A$ for all $i\in I$. Specifically, \homo{m_i}{C_i}{A} is the inclusion and $\homo{q_i}{A}{C_i}$ is the homomorphism induced by \cterm{\homo{f_{j,i}}{C_j}{C_i}}{j\in I} where $f_{j,i}$ sends $C_j$ to the core if $j\neq i$, and $f_{i,i}$ is the identity.
Next, to show that $c_i$ is maximal in $A$ for all $i\in I$:
suppose there  exists a homomorphism \homo{h}{A}{A} and a $t:A$ such that $h(t)=c_i$. Then $t$ is contained in some $C_j$. By composing 
\[\bfig
\morphism(600,0)<300,0>[A`C_i;q_i]
\morphism(300,0)<300,0>[A`A;h]
\morphism<300,0>[C_j`A;m_j]
\efig\]
and by \ref{lemma:whencoresarecovered},  we see that $i=j$ and $(q_i\comp h\comp m_i)$ is an isomorphism on $C_i$. Hence $m_i\comp (q_i\comp h\comp m_i)^{-1}\comp q_i(c_i)=t$.
Finally, if $k:A$ is maximal, then $k:C_i$ for some $i\in I$, whence $C_k$ is a reflective subinstance of $C_i$. 
\end{proof}

 Let \canrep{A}  be  the  union of all members of the multicore, where these are chosen so as to have no nulls in common, $\canrep{A}:=\displaystyle\bigcup_{E\in\multicore{A}}E$.
It is now easy to see that $\canrep{A}\equiv A$, so that $\canrep{-}$ is a representative function, in the sense of \ref{sec:semanticprelims}. 
%
%
%
\canrep{A} need not be minimal either in terms of subinstances or images.
However, an instance is sub-minimal only if it is an image of \canrep{A}, as we show by  way of the following lemma.
\begin{lemma}\label{thm:the arrow from the canonical rep}
Let $A\in \mathcal{D}$. There exists a homomorphism $m:\canrep{A}\rightarrow A$ such that $\canrep{A}\equiv m(\canrep{A})\equiv A$.
\end{lemma}
\begin{proof}
We can regard $\multicore{A}$ as a set  \cterm{C^A_k}{k\in S} for a suitable set $S$. For each $k\in S$ we have an inclusion $i_k: C_k\rightarrow A$,  and a strong surjection $s_k:A\rightarrow C_k$.
The family \cterm{i_k: C_k\rightarrow A}{k\in S} determines a homomorphism  $m:\canrep{A}\rightarrow A$. For each $k$ the inclusion  $i_k: C_k\rightarrow A$ factors through $m(\canrep{A})$ and the composite 
$C_k\to^{i_k} m(\canrep{A})\subseteq A \to^{s_k} C_k$
is the identity. Thus $\canrep{A}\equiv m(\canrep{A})\equiv A$.
%
\end{proof}
\begin{theorem}
An instance $A$ is sub-minimal only if there exists a strongly surjective homomorphism $m:\canrep{A}\rightarrow A$.
\end{theorem}
%
%
\begin{corollary}
Identifying isomorphic instances, the number of sub-minimal instances that are semantically equivalent to  $A$ is bounded by the number of (semantically equivalent) images of \canrep{A}.
\end{corollary}
\begin{remark}\label{rem:gluingmulticores}
We note that there will usually be proper semantically equivalent images of \canrep{A}. In particular, this always exists if the core of $A$ has a null in it and \multicore{A} has more than one member. The reason is that members of \multicore{A} can be `glued' along common reflective subinstances; such subinstances induce a filter which yields a semantically equivalent image of \canrep{A}. 
Observe that if \multicore{A} has a \emph{single} member, then that member is equivalent to $A$, and thus $[A]$ has a least element both in terms of subinstances, reflective subinstances, and images. 
\end{remark}


%
\begin{example}\label{ex: final return to ex 12}
Consider \ref{example:No least R-quotient}.
 $\multicore{A_2}$ consists of the two PCWA-cores $C_{k_1}$ and $C_{k_5}$. In addition to the core, $C_{k_1}$ and $C_{k_5}$ have the  reflective subinstances $V$ and $W$ in common.
\begin{center}
\scalebox{0.8}{ \begin{tabular}{ l | c c c c r }
$C_{k_1}$ &R&&&&\\ 
\hline
 $k_1$  &x & x & u & y & z \\
 $k_2$   &x & x &  x & x & z\\
   $k_3$ &x & x & x  & y & x\\
   $k_4$&x & x & x  & x & x\\  
\end{tabular}}
\quad
\scalebox{0.8}{ \begin{tabular}{ l | c c c c r }
$C_{k_5}$ &R&&&&\\
\hline

    $k_5$&v & p & p & r & s \\
    $k_6$& p& p &  p & p & s\\
    $k_7$&p & p & p  & r & p\\
   $k_8$&p & p & p  & p & p\\
\end{tabular}}
\quad 
\scalebox{0.8}{ \begin{tabular}{ l | c c c c r }
 $V$&R&&&&\\
\hline
    &x & x & x  & y & x\\
   &x & x & x  & x & x\\  
\end{tabular}}
\quad
\scalebox{0.8}{ \begin{tabular}{ l | c c c c r }
 $W$&R&&&&\\
\hline
    & p& p &  p & p & s\\
   &p & p & p  & p & p\\
\end{tabular}}
\end{center}   
%
\end{example}
The filter induced on $C_{k_1}\cup C_{k_5}=A_2$ by $V$ identifies $x$ with $p$ and $y$ with $r$. If we write out the resulting image by overwriting $x$ with $p$ and $y$ with $r$, we obtain $B_2$ of \ref{example:No least R-quotient}. It follows that $B_2$ is a semantically equivalent image of $A_2$. Similarly, from  $W$ we see that we can produce a semantically equivalent image of $A_2$ by overwriting $x$ with $p$ and $z$ with $s$. This results in $C_2$.

\subsection{Na\"{i}ve Evaluation of Queries}\label{sec:queries}

Before proceeding to the study of  minimality for OCWA* in general, we make an example remark on the use of \multicore{-} in the evaluation of queries. The motivation is, briefly, that it may be significantly cheaper to evaluate a query separately on the smaller instances in \multicore{A} than  on all of $A$. 

Recall from e.g.\ \cite{Gheerbrant:2014:NEQ:2691190.2691194} that that the \emph{certain answers} of a query $Q$ on an instance $A$ under a semantics \Rep[]{-} is the intersection of the answers obtained on \Rep[]{A}: $\mathsf{certain}(Q,A):=\bigcap\cterm{Q(I)}{I\in \Rep[]{A}}$.
The \emph{na\"{i}ve evaluation} of $Q$ on $A$ is the result of removing all tuples with  nulls from $Q(A)$. Na\"{i}ve evaluation is said to \emph{work} for $Q$ if it produces precisely the certain answers.

It is shown in \cite{Gheerbrant:2014:NEQ:2691190.2691194} that na\"{i}ve evaluation works for the class $\exists\mathbf{Pos}+\forall\mathbf{G}^{\mathrm{bool}}$ of  existential positive queries with Boolean universal guards with respect to PCWA. 
%
Here $\exists\mathbf{Pos}+\forall\mathbf{G}^{\mathrm{bool}}$ is the least class of formulas containing all atomic formulas, including equality statements, and closed under conjunction; disjunction; existential quantification; and the following rule: if $\alpha$ is an atomic formula,  $\phi$ a formula in $\exists\mathbf{Pos}+\forall\mathbf{G}^{\mathrm{bool}}$, and \x\ a list of distinct variables containing all free variables in both $\alpha$ and $\phi$, then $\forall \x (\alpha\rightarrow \phi)$ is a formula in $\exists\mathbf{Pos}+\forall\mathbf{G}^{\mathrm{bool}}$.

%
\begin{theorem}
Let   $Q$ be a query of arity $n$ in $\exists\mathbf{Pos}+\forall\mathbf{G}^{\mathrm{bool}}$. Then $\mathsf{certain}(Q,A)=\bigcap\cterm{Q(E)}{E\in \multicore{A}}\cap \mathsf{Const}^n$.
\end{theorem}
\begin{proof}
The inclusion from left to right follows from the fact that na\"{i}ve evaluation works for $Q$, that $Q$ is preserved under strong surjections, and that each $E\in\multicore{A}$ is an image of $A$. For the inclusion from right to left, it is sufficient to show that if $\a$ is a tuple of constants in  $\bigcap\cterm{Q(E)}{E\in \multicore{A}}$ then $\a\in Q(I)$ for all $I\in \overline{\cup_{E\in \multicore{A}}\Rep[CWA]{E}}^{\cup}$. But this is a straightforward modification of the proof that formulas in $\exists\mathbf{Pos}+\forall\mathbf{G}^{\mathrm{bool}}$ are preserved under unions of strong surjections (Lemma 10.12) in  \cite{Gheerbrant:2014:NEQ:2691190.2691194}.
\end{proof}

\section{Minimality in OCWA*}
\label{sec:minforocwa}

We return now to OCWA* in general and apply our results from the special case of the previous section. 
Recall from \ref{sec:EqandMinforOCWA} that in order to determine whether $A[\X]\leq_{\mathsf{OCWA}*} B[\Y]$ we have to look for an RCN-cover from $B[\Y]$ to $A^{\small \mathrm{length(\Y)+1}}[\X]$.
The reason is that RCN-covers do not compose; it is insufficient just to know that we have an RCN-cover from $B[\Y]$ to $A[\X]$.
This fact complicates the study of minimality for OCWA*.
 However, note that   if there exists an RCN-cover from  $B[\Y]$ to $A[\X]$ which sends the closed nulls \Y\  to closed terms in $A[\X]$---i.e.\ either to \X\ or to constants---then, because such covers do compose, we have $A[\X]\leq_{\mathsf{OCWA}*} B[\Y]$. 
But, on the face of it, we cannot restrict to such `closedness-preserving' covers. 
Consider the following example.
%
%
%
%
\begin{example}\label{ex:redundancy}The following annotated instances are all semantically equivalent.
\small
\begin{align*}
A[V,W] & = \set{R(x,y), R(V, W)} \quad A[V,w] = \set{R(x,y), R(V, w)} \quad A[v,w] = \set{R(x,y), R(v, w)}\\
A[v,W] & = \set{R(x,y), R(v, W)} \quad  B[]  = \set{R(x,y)}	
\end{align*}
\end{example}
Although $A[V,W]$ and $B[]$ are equivalent, there is no RCN-cover from $A[V,W]$ to $B[]$ that satisfies the restriction that  closed nulls should be sent  to closed terms, since everything in $B[]$ is open. It is the (ordinary) RCN-cover to $B^3[]$ that witnesses that $B[]\leq_{\mathsf{OCWA}*} A[V,W]$.
%
%
Nevertheless, \ref{ex:redundancy} hints at a solution to this; the problem with $A[V,W]$, it can be said, is that it has closed nulls that could equivalently have been annotated as open. Once we re-annotate to $A[v,w]$, the equivalence with $B[]$ \emph{is} witnessed by a cover to $B[]$.
 We show in this section that if we restrict to annotated instances where no closed null can be equivalently replaced by an open null, then any semantic equivalence is witnessed by RCN-covers which preserve closed nulls. These closed nulls can then, essentially, be treated as constants, so that the results of \ref{sec:minforpcwa} can be applied.

We note, first, when   an annotated instance is semantically equivalent to an instance without closed nulls. Our fixed semantics in this section is OCWA*.

\begin{proposition}\label{prop:semiopencharacterization} Let $A[\X]$ be an annotated instance. The following are equivalent:
\begin{enumerate}
\item $A[\X]$ is semantically equivalent to an instance $B[]$ in which all nulls are open.
\item $\Rep[]{A[\X]}$ is closed under unions.
\item $A[\X]$ is semantically equivalent to the  instance $A[\x]$ obtained by changing the annotation of $A[\X]$ so that all nulls are open.
\end{enumerate}
\end{proposition}
\begin{proof}
For 2.\ $\Rightarrow$ 3., note  
that $\Rep[]{A[\X]}\subseteq \Rep[]{A[\x]}$ is clear; and 
it is also clear that $\Rep[CWA]{A[\x]}\subseteq \Rep[]{A[\X]}$. But then, since \Rep[]{A[\X]} is closed under unions, $\Rep[]{A[\x]}\subseteq \Rep[]{A[\X]}$.
\end{proof}

\begin{definition}
Let $A[\X,\BF{Y}]$ be an annotated  instance. We say that $\X$ is \emph{annotation redundant} (relative to \BF{Y}) if 
$ A[\X,\BF{Y}]\equiv A[\x,\BF{Y}]$
i.e.\ if changing the annotation of \X\ to   ``open'' yields an equivalent instance. We say that an annotated incomplete instance is \emph{annotation minimal} is no subset of its closed nulls are annotation redundant (with respect to the rest).
\end{definition}

%
%
\begin{lemma}\label{lemma:redchar}
Let $A[\X,\BF{Y}]$ be an annotated incomplete instance and \cc\ a list of the same length as \Y\ of distinct constants not occurring in $A$ . Then $\X$ is redundant with respect to \Y if and only if for all finite lists of instances $I_1,\ldots, I_k\in\Rep[]{A[\X, \cc]}$ it is the case that $I_1 \cup \ldots \cup I_k\in\Rep[]{A[\X, \Y]}$.
\end{lemma}
\begin{proof}
\emph{If}: We must show that $A[\X,\BF{Y}]\equiv A[\x,\BF{Y}]$. Let $m\geq 1$ be given, and consider $ A^m[\x,\BF{Y}]$. Let $J$ be a freeze of  $ A^m[\x,\BF{Y}]$ where \Y\ is replaced by \cc\ and the other nulls by fresh constants. Then $J=I_1\cup\ldots\cup I_m$ where $I_1,\ldots, I_m\in \Rep[]{A[\X,\cc]}$.  So $J\in A[\X,\BF{Y}]$, by assumption, and then $A[\X,\BF{Y}]\equiv A[\x,\BF{Y}]$ by \ref{thm:cover-chara}.  

\emph{Only if}: If $I_1,\ldots, I_k\in\Rep[]{A[\X, \cc]}$ then $I_1,\ldots, I_k\in\Rep[]{A[\x, \cc]}$, and then,  since the latter is closed under unions,  $I_1 \cup \ldots \cup I_k\in\Rep[]{A[\x, \cc}\subseteq\Rep[]{A[\x, \Y]}=\Rep[]{A[\X,\Y]}$.
\end{proof}


The following theorem displays the main property of annotation-minimal instances. The proof is rather long
 and is omitted for reasons of space. 
%
%
\begin{theorem}\label{prop:mainOCWAprop}
Let $A[\X]$ and $B[\Y]$ be two annotation-minimal instances such that $A[\X]\equiv B[\Y]$. Then for all strong surjections $f:A^{\infty}[\X]\twoheadrightarrow B^{\infty}[\Y]$ it is the case that $f$ restricts to a bijection $f\upharpoonright_{\X}:\X\rightarrow \Y$ on the sets of closed nulls.    
\end{theorem}
\begin{corollary}
Let $A[\X]$ and $B[\Y]$ be two annotation-minimal instances.  Then $A[\X]\equiv B[\Y]$
if and only if there exists 
RCN-covers
\cterm{f_i:A[\X]\rightarrow B[\Y]}{1\leq i\leq n} and 
 \cterm{g_j:B[\Y]\rightarrow A[\X]}{1\leq j\leq m}
such that $f_i$ restricts to a bijection $f_i\upharpoonright_{\X}:\X\rightarrow\Y$ and $g_j$ to a bijection  $g_j\upharpoonright_{\Y}:\Y\rightarrow\X$.
\end{corollary}

\begin{corollary}\label{cor:substituteconstantscorollary}
Let $A[\X]$ and $B[\Y]$ be two annotation-minimal instances.  Then $A[\X]\equiv B[\Y]$ if and only if $\X$ is of the same length as $\Y$ and there exists injective functions $f:\X\rightarrow \mathsf{Const}$ and $g:\Y\rightarrow \mathsf{Const}$ such that $A[f(\X)/\X]\equiv_{\mathsf{PCWA}} B[g(\Y)/\Y]$, where $f(\X)$ and $g(\Y)$ are disjoint from the constants in $A[\X]$ and $B[\Y]$.
\end{corollary}
That is to say, $A[\X]$ and $B[\Y]$ are equivalent if there is a way to `freeze' the closed nulls so that they become PCWA-equivalent. 

Now, let $A[\X]$  be an annotation-minimal instance.  Let \cc\ be a list of fresh constants, of the same length as \X. Then we can compute $\multicore{A[\cc]}=\{ E_1,\ldots, E_n \}$ and $\canrep{A[\cc]}=E_1\cup\ldots \cup E_n$, as in \ref{sec:minforpcwa}, and then substitute \X\ back in for \cc. This yields a set $\multicore{A[\X]}:=\{E_1[\X], \ldots, E_n[\X]\}$ of annotated instances  and an annotated instance $\canrep{A}[\X]:=E_1[\X]\cup\ldots \cup E_n[\X]$.
Since $\canrep{A}[\X]$ is semantically equivalent to $A[\X]$ and has the same (number of) closed nulls, $\canrep{A}[\X]$ is annotation-minimal.
Thus we have a function $\multicore{-}$ from annotation minimal annotated instances to finite sets of annotated instances, and \canrep{-} from annotation minimal  instances to annotation minimal instances.   
As in \ref{sec:minforpcwa}, we identify \multicore{A[\X]} and  \multicore{B[\Y]} if they `are the same up to renaming of nulls', but in the presence of closed nulls we have to add a condition to what this means: we say that $\multicore{A[\X]}=\multicore{B[\Y]}$ if there is a bijection of sets $F$ between them; an isomorphism $f_E:E\rightarrow F(E)$ for each $E\in\multicore{A[\X]}$; and for all  $E,E'\in\multicore{A[\X]}$, the homomorphisms $f_E$ and $f_{E'}$ restrict to one and the same bijection of sets $\X\rightarrow \Y$.    
We now have:

\begin{theorem}\label{thm:ocwa multicore}
Let  $A[\X]$, $B[\Y]$   be an annotation-minimal instances. Then:
\begin{enumerate}
\item  $A[\X]\equiv B[\Y]$ if and only if $\multicore{A[\X]}=\multicore{B[\Y]}$;
\item  $A[\X]\equiv \canrep{A[\X]}$, and \canrep{A[\X]} is annotation-minimal; 
\item if $E[\X]\in \multicore{A[\X]}$ and $A[\X]\equiv B[\Y]$ then $E[\X]$ is a reflective subinstance of $B[\Y]$ (up to annotation-preserving isomorphism); and
\item  if $A[\X]\equiv B[\Y]$ then $B[\Y]$ is sub-minimal only if there is a strongly surjective homomorphism\\ $\canrep{A}[\X]\twoheadrightarrow B[\Y]$ restricting to a bijection $\X\rightarrow\Y$. 
\end{enumerate} 
\end{theorem}
Accordingly, $\multicore{-}$ is representative in the sense of  (1) and (2) and minimal in the sense of (3). \canrep{-}  bounds the number of sub-minimal equivalent instances by (4).  \multicore{-} (and \canrep{-}) can be extended to all annotated instances by first choosing an equivalent annotation minimal instance and then applying \multicore{-}, and (1)  ensures that the result does not depend on the choice.

\section{Discussion and Conclusion}
\label{sec: conclusion}

In this work we study the problems of implication, equivalence, and minimality (and consequently cores) in mixed open and closed worlds. 
These problems have particular importance in the context of date exchange
and remain open for several variants of mixed worlds.
In particular, we adress these problems for the Closed Powerset semantics and the OCWA semantics. 
To this end, we define a novel semantics for mixed worlds that we called OCWA*
and subsumes both Closed Powerset and OCWA.
Our semantics is introduced with the help of homomorphic covers and it is characterised in terms of such covers.  
For the minimization problem we presented negative results for several common notions of minimality. 
Then, we showed that one can find cores using a different notion of minimality.

Observe that homomorphic covers have been already used in several related contexts. 
In \cite{Grahne:2015:RED:2745754.2745770}, Grahne et al.\ uses homomorphic covers in the context of source instance recovery in data exchange.
In \cite{Chaudhuri:1993:ORC:153850.153856}, Chaudhuri and Vardi give the existence of a cover as a sufficient condition for conjunctive query containment under bag semantics. In \cite{DBLP:journals/tods/KostylevRS14}, Kostylev et al.~use various notions of cover to study annotated query containment. On the other hand, Knauer and Ueckerdt \cite{Knauer2016745} apply this notion to coverage relations between graphs.

In our opinion
several more data management scenarios can benefit from the concept of homomorphic cover and the machinery that we have developed for it.
For instance, two conjunctive queries whose relational structures cover each other retrieve the same tuples from every relation of any database instance, a fact of potential relevance in e.g.\ data privacy settings. In the field of constraint programming, this property is closely connected to the notion of a minimal constraint network \cite{DBLP:journals/ai/Gottlob12}, and may have applications there. For another example, treating one conjunctive query as a view, it can be used to completely rewrite another if there exists a cover from the view (cf.~\cite{DBLP:conf/pods/LevyMSS95}). Thus in this setting, cover-equivalence corresponds to mutual complete rewritability. 



\paragraph{Acknowledgement:}

This work was partially supported by the SIRIUS Centre, Norwegian Research Council project number 237898 and project number 230525.

\bibliographystyle{plain}
\bibliography{optique,mainrefs-thesis,pers-refs}

\newpage

\section*{Appendix}
\subsection*{Proofs, examples, and remarks omitted for reasons of space}

\begin{appthm}{\bf \ref{proposition: Libkinrewrite}}
Let $A$ be an annotated instance. Then there exists an annotated instance $A'$ on normal form such that $\Rep[OCWA(LS)]{A}=\Rep[OCWA(LS)]{A'}$.
\end{appthm}
\begin{proof}
Say that each occurrence in $A$ of an open constant or an open null that occurs elsewhere in $A$ as closed is a \emph{flagged term occurrence}. Likewise, say that each atom in which there is a flagged term occurrence is a  \emph{flagged atom}.   
Then  $A'$ is constructed  from $A$ by replacing each flagged atom $R(\mathbf{t})$ by that same atom which each flagged term occurrence re-annotated as closed, together with a new atom $R(\mathbf{ t'})$ obtained by replacing each flagged term occurrence by a fresh open null. Unflagged atoms remain the same.    
There is  an evident strong surjection $e:A'\rightarrow A$ and an inclusion $m:A\rightarrow A'$ such that $e\circ m$ is the identity on $A$.

Suppose $I\in \Rep[OCWA(LS)]{A}$. Then there exists a homomorphism $h:A\rightarrow I$ such  that 
for every $R(\mathbf{ v})$ in $I$
	there exists a 
	$R(\mathbf{ t})$ in $A$ such that 
	 $h(\mathbf{ t})$ and $\mathbf{ v}$ 
	agree on all positions annotated as closed in $\mathbf{ t}$. But then, if $\mathbf{ t}$ is not flagged, $\mathbf{ v}$ agrees with $h\circ e (m(\mathbf{ t}))$ on all positions annotated as closed, and if $\mathbf{ t}$ is flagged, $\mathbf{ v}$ agrees with $h\circ e(\mathbf{ t'})$ on all positions annotated as closed.   So $I\in \Rep[OCWA(LS)]{A'}$.
	 Conversely, assume that $I\in \Rep[OCWA(LS)]{A'}$, witnessed by a homomorphism $g:A'\rightarrow I$, and suppose that  $\mathbf{ v}$  agrees with $g(\mathbf{ u})$ on all closed positions. If $\mathbf{u}=m(\mathbf{t})$ then   $\mathbf{ v}$ agrees with $g\circ m(\mathbf{ t})$ on all closed positions in $\mathbf{t}$. If not, then $\mathbf{ u} = \mathbf{ t}' $ for some flagged $R(\mathbf{ t})$ in $A$, in which case $\mathbf{v}$  agrees with $g\circ m(\mathbf{ t})$ on all closed positions.
\end{proof}

\noindent Example regarding maximal tuples, omitted from text:
\begin{example}\label{ex:maximaltuples}
Consider $A_2$ from \ref{example:No least R-quotient} (with names of atoms added)
\begin{center}
\scalebox{0.8}{\begin{tabular}{ l | c c c c r }
$A_2$ &R&&&&\\
\hline
 $k_1$  &x & x & u & y & z \\
 $k_2$   &x & x &  x & x & z\\
   $k_3$ &x & x & x  & y & x\\
   $k_4$&x & x & x  & x & x\\
    $k_5$&v & p & p & r & s \\
    $k_6$& p& p &  p & p & s\\
    $k_7$&p & p & p  & r & p\\
   $k_8$&p & p & p  & p & p\\
\end{tabular}}
\end{center}
The only maximal atoms are $k_1$ and $k_5$. We have 
\begin{center}
\scalebox{0.8}{\begin{tabular}{ l | c c c c r }
$C_{k_1}$ &R&&&&\\
\hline
 $k_1$  &x & x & u & y & z \\
 $k_2$   &x & x &  x & x & z\\
   $k_3$ &x & x & x  & y & x\\
   $k_4$&x & x & x  & x & x\\  
\end{tabular}}
\quad
\scalebox{0.8}{\begin{tabular}{ l | c c c c r }
$C_{k_5}$ &R&&&&\\
\hline

    $k_5$&v & p & p & r & s \\
    $k_6$& p& p &  p & p & s\\
    $k_7$&p & p & p  & r & p\\
   $k_8$&p & p & p  & p & p\\
\end{tabular}}
\end{center}   
\end{example}
%

%
%
%
%
%

\begin{appthm}{\bf \ref{rem:gluingmulticores}}
We observe that there will usually be proper semantically images of \canrep{A}. In particular, this always exists if the core of $A$ has a null in it and \multicore{A} has more than one member, so that the core occurs more than once in \canrep{A}. 

Specifically, and  more generally, let $A$ and $B$ be   instances, and assume for convenience  that they have no nulls in common. 
Suppose there is a strict reflective subinstance $C_A$ of $A$ and  a strict reflective subinstance $C_B$ of $B$ so that $C_A\cong C_B$. 
%
Fix an isomorphism \homo{f}{C_A}{C_B}. $f$ induces a filter on $A\cup B$ as the symmetric and reflexive closure of the relation $x\simeq_f y \Leftrightarrow f(x)=y$. We obtain a strong surjection  $\homo{q}{A\cup B}{A\cup B/_{\simeq_f}}$ by identifying the nulls related by $\simeq_f$. But we also have strong surjections $A\cup B/_{\simeq_f}\rightarrow A$ and     $A\cup B/_{\simeq_f}\rightarrow B$, so $A\cup B$ and $A\cup B/_{\simeq_f}$ are semantically equivalent.
\end{appthm}

\begin{lemma}
\label{lemma:mainocwalemma}
Let $A[\X]$  be an annotation-minimal instance, and let $f:A^{\infty}[\X]\twoheadrightarrow A^{\infty}[\X]$ be strongly surjective. Then $f$ fixes \X\ setwise.
\end{lemma} 
%
\begin{proof}

First, suppose for contradiction that it is not the case that $f(\X)\subseteq \X$.

Write \X\ as \Y,\Z, where $f(\Y)$ are not in \X\ but $f(\Z)$ are. 

Let  \cc\  be a list of distinct constants not occurring in $A$ of the same length as \Z\ and $I_1,\ldots, I_k\in\Rep{A[\Y, \cc/\Z]}{}$; witnessed by strong surjections
\[s_i:A^{\infty}[\Y,\Z]\twoheadrightarrow I_i\]
where $s_i(\Z)=\cc$ for $1\leq i\leq k$.

Recall that  $\nabla :A^{\infty}[\X]\rightarrow A[\X]$ is the homomorphism which identifies duplicates of open nulls. 

Let $g:A[\X]\rightarrow \overline{A[\X]}$ be a (fresh) freeze of $A[\X]$. Write $\mathbf{t}:=g \circ\nabla\circ f(\Y)$, 
\[g\circ \nabla\circ f:A^{\infty}[\X]\twoheadrightarrow A^{\infty}[\X]\twoheadrightarrow A[\X]\twoheadrightarrow \overline{A[\X]} \]
Note that $\mathbf{t}$ and $g(\X)$ have no constants in common. 

Since $g(\X)$ is a list of distinct constants, $g(\X)\mapsto s_i(\X)$ defines a partial function from constants to constants, and we can extent this to a full function  $h_i:\mathsf{Const}\rightarrow\mathsf{Const}$  by setting $h_i(a)=a$ for all constants not in $g(\X)$. This defines a complete instance 
\[J_i:=h_i(\overline{A[\X]})\]
as the image of $\overline{A[\X]}$ under $h_i$. We write $h_i:\overline{A[\X]}\twoheadrightarrow J_i$ also for the strongly surjective \emph{structure} homomorphism. 

 
Note that $h_i(\mathbf{t})=\mathbf{t}$ (and that the composite 
\[h_i\circ g\circ \nabla\circ f:A^{\infty}[\X]\twoheadrightarrow A^{\infty}[\X]\twoheadrightarrow A[\X]\twoheadrightarrow \overline{A[\X]}\twoheadrightarrow J_i \]
is a database homomorphism).

For $1\leq i \leq k$,  we have that $h_i\circ g\circ \nabla(\X)=s_i(\X)$ so that $I_i\cup J_i\in\Rep[]{A[s_i(\X)]}= \Rep[]{A[s_i(\Y), \cc/\Z]}$.
 
Now, we can write $A^{\infty}[\X]$ as $A^{p}[\X]\oplus A^{\infty}[\X]$ 
\[f:A^{\infty}[\X]\twoheadrightarrow A^{p}[\X]\oplus A^{\infty}[\X]\]
so that $f(\Y)\in A^{p}[\X]$. Then
\[(h_i\circ g\circ \nabla ,s_i)\circ f:A^{\infty}[\X]\twoheadrightarrow A^{p}[\X]\oplus A^{\infty}[\X]\twoheadrightarrow J_i \cup I_i\]
is well-defined,  and since 
 \begin{align*}
( (h_i\circ g\circ \nabla ,s_i)\circ f)(\Y) &=  (h_i\circ g\circ \nabla \circ f)(\Y) \\
                                                                                         &= \mathbf{t}                                                                   
 \end{align*}
 and
 \begin{align*}
 ( (h_i\circ g\circ \nabla ,s_i)\circ f)(\Z) &= (s_i(f(\Z)) 
 \end{align*}
 we have that $J_i\cup I_i\in \Rep[]{A[\mathbf{t}/\Y, s_i(f(\Z))/\Z]}$.
 
 But $ s_i(f(\Z))= s_j(f(\Z))$ for all $1\leq i,j \leq k$, so 
 \[\bigcup_{i=1}^{k} (J_i\cup I_i)\in \Rep[]{A[\mathbf{t}/\Y, s_1(f(\Z))/\Z]}\]
 
 Finally, the map $\pi_1: A[\X]\rightarrow A^{\infty}[\X]$  induces a structure homomorphism $r_i:J_i\rightarrow I_i$ which fixes the constants in $A$ and $s_i(\X)$ pointwise. Thus we have a strongly surjective structure homomorphism  
 \[\bigcup_{i=1}^{k} (J_i\cup I_i)\twoheadrightarrow \bigcup_{i=1}^{k} (I_i)\]
 which fixes the constants in $A$ pointwise. Whence 
 \[\bigcup_{i=1}^{k} (I_i)\in \Rep[]{A[\Y, \Z]}\]
 since this set is closed under such surjections. Thus $\Y$ is redundant by \ref{lemma:redchar}, contrary to assumption. We conclude that 
 \[f(\X)\subseteq \X\]
 
Now, suppose that $f(\X)$ is a proper subset of \X. We write $\X=\Y,\Z$ where $f(\X)=\Z$.  

Let  \cc\  be a list of distinct constants not occurring in $A$ of the same length as \Z\ and $I_1,\ldots, I_k\in\Rep[]{A[\Y, \cc/\Z]}$; witnessed by strong surjections
\[s_i:A^{\infty}[\Y,\Z]\twoheadrightarrow I_i\]
where $s_i(\Z)=\cc$ for $1\leq i\leq k$. Then
 \[\bigcup_{i=1}^{k} (I_i)\in \Rep{A[s_1(f(\X))/\X]}{}\]
 Whence, again, \Y\ is redundant by \ref{lemma:redchar}, contrary to assumption.
 
 We conclude that $f$ fixes \X\ setwise. 
\end{proof}

\begin{appthm}{\bf \ref{prop:mainOCWAprop}}
Let $A[\X]$ and $B[\Y]$ be two annotation-minimal instances such that $A[\X]\equiv B[\Y]$. Then for all strong surjections $f:A^{\infty}[\X]\twoheadrightarrow B^{\infty}[\Y]$ it is the case that $f$ restricts to a bijection $f\upharpoonright_{\X}:\X\rightarrow \Y$ on the sets of closed nulls.    
\end{appthm}
\begin{proof}
Let strongly surjective homomorphisms 
\[f:A^{\infty}[\X]\leftrightarrows B^{\infty}[\Y]:e\]
 be given.
  Write $\X=\X_1,\X_2$, $\Y=\Y_1,\Y_2$ where a closed null is in $\X_2$ if $f$ sends it to a closed null, i.e.\ to a $Y$ in $\Y$, and $\Y_2$ is the image of $\X_2$. Assume for contradiction that $\X_1$ is non-empty.

   By \ref{lemma:mainocwalemma}, $e\circ f$ and $f\circ e$ has to fix \X\ and \Y, respectively, setwise. Therefore:  $f\upharpoonright _{\X_2}:\X_2\rightarrow \Y_2$ must be injective; $e(\Y_2)\subseteq \X_2$; and     $e\upharpoonright _{\Y_2}:\Y_2\rightarrow \X_2$ must be injective as well. In summary,  $f\upharpoonright _{\X_2}\circ e\upharpoonright _{\Y_2}$ and $e\upharpoonright _{\Y_2}\circ f\upharpoonright _{\X_2}$ are permutations of $\Y_2$ and $\X_2$, respectively, and $e(\Y_1)$ are open nulls or constants in $A^{\infty}[\X]$. 

Let  \cc\  be a list of distinct constants not occurring in $A$ of the same length as $\Y_2$ and $I_1,\ldots, I_k\in\Rep[]{B[\Y_1, \cc/\Y_2]}$; witnessed by strong surjections
\[s_i:B^{\infty}[\Y_1,\Y_2]\twoheadrightarrow I_i\]
where $s_i(\Y_2)=\cc$ for $1\leq i\leq k$.

Let $g:B[\Y_1,\Y_2]\rightarrow \overline{B[\Y_1,\Y_2]}$ be a (fresh) freeze of $B[\Y_1,\Y_2]$. Write $\mathbf{t}:=g \circ\nabla\circ f(\X_1)$.  
\[g\circ \nabla\circ f:A^{\infty}[\X]\twoheadrightarrow B^{\infty}[\Y]\twoheadrightarrow B[\Y]\twoheadrightarrow \underline{B[\Y]} \]
Note that $\mathbf{t}$ and $g(\Y)$ have no constants in common.

Since $g(\Y)$ is a distinct list of constants, $g(\Y)\mapsto s_i(\Y)$ is a partial function of constants, for each  $1\leq i \leq k$. Extend this to a function $h_i:\mathsf{Const}\rightarrow \mathsf{Const}$ by setting $h_i(a)=a$ for  all $a$ not in \Y. Set  $J_i:=h_i(\overline{B[\Y]})$.
 %
Note that $h_i(\mathbf{t})=\mathbf{t}$.

\[h_i\circ g\circ \nabla\circ f:A^{\infty}[\X]\twoheadrightarrow B^{\infty}[\Y]\twoheadrightarrow B[\Y]\twoheadrightarrow \overline{B[\Y]}\twoheadrightarrow J_i \]

For $1\leq i \leq k$,  we have that $h_i\circ g\circ \nabla(\Y)=s_i(\Y)$ so that $I_i\cup J_i\in \Rep[]{B[s_i(\Y)]}$.
 
Now, we can write $B^{\infty}[\Y]$ as $B^{n}[\Y]+_{\Y} B^{\infty}[\Y]$ and 
\[f:A^{\infty}[\X]\twoheadrightarrow B^{n}[\Y]+_{\Y} B^{\infty}[\Y]\]
so that $f(\X_1)\in B^{n}[\Y]$. Then we have
\[(h_i\circ g\circ \nabla ,s_i)\circ f:A^{\infty}[\X]\twoheadrightarrow B^{p}[\Y]+_{\Y} B^{\infty}[\Y]\twoheadrightarrow J_i \cup I_i\]
 and since 
 \begin{align*}
( (h_i\circ g\circ \nabla ,s_i)\circ f)(\X_1) &=  (h_i\circ g\circ \nabla\circ f)(\X_1) 
                                                                                         = \mathbf{t}                                                                   
 \end{align*}
 and
 \begin{align*}
 ( (h_i\circ g\circ \nabla ,s_i)\circ f)(\X_2) &= (s_i(f(\X_2)) 
 \end{align*}
 we have that $J_i\cup I_i\in \Rep[]{A[\mathbf{t}/\X_1, s_i(f(\X_2))/\X_2]}$.
 
 But $ s_i(f(\X_2))= s_j(f(\X_2))$ for all $1\leq i,j \leq k$, so 
 \[\bigcup_{i=1}^{k} (J_i\cup I_i)\in \Rep[]{A[\mathbf{t}/\X_1, s_1(f(\X_2))/\X_2]}{}\]
 
 Finally, the map $\pi_1: B[\Y]\rightarrow B^{\infty}[\Y]$  induces a structure homomorphism $r_i:J_i\rightarrow I_i$ which fixes the constants in $B$ and $s_i(\Y)$ pointwise. $A$ has the same constants as $B$. Thus we have a strongly surjective structure homomorphism 
 \[\bigcup_{i=1}^{k} (J_i\cup I_i)\twoheadrightarrow \bigcup_{i=1}^{k} (I_i)\]
 which fixes the constants in $A$ pointwise. Whence 
 \[\bigcup_{i=1}^{k} (I_i)\in \Rep[]{A[\X]}\]
 since this set is closed under such surjections. Thus $\X_1$ is redundant, contrary to the assumption that $A[\X]$ is annotation minimal. We conclude that $\X_2=\X$.  By symmetry,  $\Y_2=\Y$.  

\end{proof}

\end{document}